\documentclass[reprint,pre,amsmath,amssymb,aps,nofootinbib]{revtex4-2}

\usepackage{graphicx}
\usepackage{bm} 

\usepackage[T1]{fontenc} 
\usepackage{textcomp} 
\usepackage{gensymb} 

\usepackage[english]{babel}

\usepackage{letltxmacro}

\LetLtxMacro{\ORIGselectlanguage}{\selectlanguage}
\makeatletter
\DeclareRobustCommand{\selectlanguage}[1]{%
  \@ifundefined{alias@\string#1}
    {\ORIGselectlanguage{#1}}
    {\begingroup\edef\x{\endgroup
       \noexpand\ORIGselectlanguage{\@nameuse{alias@#1}}}\x}
}
\newcommand{\definelanguagealias}[2]{%
  \@namedef{alias@#1}{#2}%
}
\makeatother

\definelanguagealias{en}{english}

\begin{document}

\title{Negative pressure and cavitation dynamics in plant-like structures}

\author{Olivier Vincent}
\email{olivier.vincent@cnrs.fr}
\affiliation{CNRS, Univ. Lyon, Univ. Claude Bernard Lyon 1, Institut Lumi\`{e}re mati\`{e}re, F-69622, Villeurbanne, France}

\begin{abstract}

It is well known that a solid (e.g. wood or rubber) can be put in a tensile stress by pulling on it.
Once a critical stress is overcome, the solid breaks, leaving an empty space.
Similarly, due to internal cohesion, a liquid can withstand tension (i.e. negative pressure), up to a critical point where a large bubble spontaneously forms, releasing the tension and leaving a void (the bubble).
This process is known as cavitation.
While water at negative pressure is metastable, such a state can be long-lived.
In fact, water under tension is found routinely in the plant kingdom, as a direct effect of dehydration, e.g. by evaporation.
In this chapter, we provide a brief overview of occurrences of water stress and cavitation in plants, then use a simple thermodynamic and fluid mechanical framework to describe the basic physics of water stress and cavitation.
We focus specifically on situations close to those in plants, that is water at negative pressure nested within a structure that is solid, but porous and potentially deformable.
We also discuss insights from these simple models as well as from experiments with artificial structures mimicking some essential aspects of the structures found within plants.

\end{abstract}

\maketitle

\tableofcontents

\newpage

\section*{Introduction}

\subsection*{Negative pressure}

\begin{figure}[htbp]
  \begin{center}
  \includegraphics[scale=0.55]{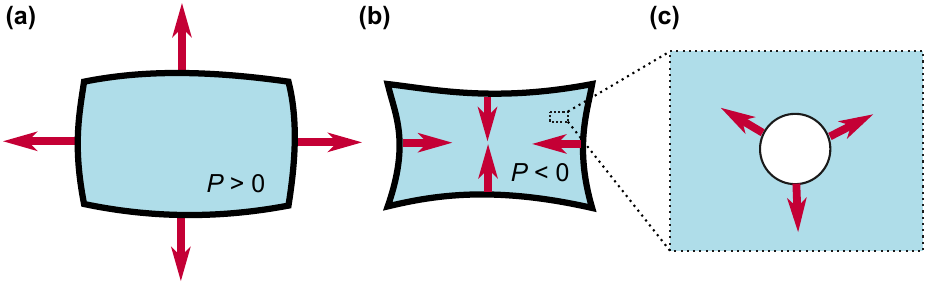}
  \caption{\small
  (a) A liquid at positive pressure exerts an outward force on the walls of a container, while
  (b) a liquid at negative pressure is in a tensile state and exerts an inward force on the walls. The walls are represented as being deformable, to illustrate the force applied by the fluid.
  (c) The tensile state as in (b) not only "pulls" on the walls but also on other interfaces; in particular, this tends to make sufficiently large bubbles grow, resulting in cavitation.
  }
\label{fig:NegativePressure}
\end{center}
\end{figure}

Liquids, similarly to solids, have internal cohesion.
The individual molecules have attractive interaction and a restoring force brings them back together if one tries to pull them apart.
Due to this internal cohesion, liquids can sustain tensile stress, and can thus be put in a state where their pressure is absolutely negative.
Going back to the definition of pressure as a force per unit area, a state of \emph{negative pressure} means that the force exerted by the fluid on a surface is directed towards the liquid and not towards the surface as it is for $P>0$ (see Figure \ref{fig:NegativePressure}a-b).
Note that because pressure in a liquid is isotropic, the tensile stress applies in all directions, while a solid can develop anisotropic stresses, e.g. with tensile stress only in one direction and compressive stresses in the other directions.

A useful analogy is that of a chain of springs: if one stretches the chain, each individual spring gets stretched and starts pulling on its neighbors; a person holding the chain would feel a force directed towards the chain.
Similarly, a liquid at negative pressure is stretched: intermolecular distances are larger than in ambient conditions, which results in a global attractive force between molecules, but also between the molecules and the container they are in.
Thus, while a liquid at positive pressure pushes on the walls of a container, a liquid at negative pressure pulls on them (see Figure \ref{fig:NegativePressure}a-b).
Obviously, the state of negative pressure can be maintained only if adhesion between the walls and the liquid is good.
In the case of water, this means that the walls of the container have to be somewhat hydrophilic.
Because of this stretched, tensile state, a liquid at negative pressure is also called \emph{under tension}, \emph{under stress}, or simply \emph{stretched}.
This stretched state is metastable and can relax by cavitation (see below).

It is useful to note here already that it is not only the sign of the pressure that matters in cavitation phenomena, but its value compared to the \emph{saturation vapor pressure}, $P_\mathrm{sat}$.
In particular, a liquid at positive pressure but with $P < P_\mathrm{sat}$ is also metastable and potentially subject to cavitation, and crossing the value $P=0$ for the liquid does not have particular consequences.
Thus, the term "negative pressure" in the following should be broadly interpreted as $P - P_\mathrm{sat} < 0$, and "stretched" as meaning "more stretched than at saturation".
For the phenomena we discuss in this chapter, this distinction remains anecdotal, because $P_\mathrm{sat}$ is several orders of magnitude below the typical negative pressures of interest ($-P \gg P_\mathrm{sat}$), so that considering $P_\mathrm{sat} \simeq 0$ does not make a significant difference.

\subsection*{Cavitation}

In a state of negative pressure, the liquid tends to "pull" on any interface it is in contact with, and in particular it will tend to make a bubble grow (see Figure \ref{fig:NegativePressure}c), if the bubble is sufficiently large (see below)\footnote{Following the remark about $P < P_\mathrm{sat}$ above, bubble growth also happens if $0 < P < P_\mathrm{sat}$, because the bubble fills with water vapor at $P_\mathrm{sat}$, and the pressure in the liquid is below $P_\mathrm{sat}$, so that there is a net outwards driving force on the bubble.}.
Bubble growth only stops when the negative pressure vanishes, which happens when bubble expansion has allowed the stretched liquid to contract back into a non-stretched state.
This process of bubble expansion associated with the disappearance of negative pressure and the relaxation of metastability is what we define as \emph{cavitation}.

Following this definition, cavitation requires a \emph{germ}, or \emph{nucleus}, i.e. a microscopic bubble in the liquid that gets stretched by the tensile forces in the liquid into a macroscopic bubble.
While such germs can pre-exist (e.g. bubbles trapped in walls, or stabilized free-floating bubbles), they can also form spontaneously by activation from thermal fluctuations.

It may seem at first sight that liquid at negative pressure should be unstable, because cavitation would always occur from existing or activated germs.
However, another driving force counteracts the stretching effect of the negative pressure on bubbles: surface tension, which accounts for the energetic cost of having a liquid-vapor interface and tends to make it as small as possible, i.e. make the bubble collapse.
Because the surface-to-volume ratio increases when bubble size decreases, surface tension dominates for small bubbles.
In other words, germs have to be above a \emph{critical size} to lead to cavitation.
Since large germs are less likely to pre-exist or to form spontaneously than small germs, a liquid at negative pressure is in fact not unstable, but metastable and can be long-lived.
In practice, cavitation occurs only if the magnitude of the negative pressure exceeds some threshold that greatly depends on the microscopic mechanism at the origin of cavitation.

Cavitation is a particular type of \emph{nucleation} due to pressure changes. Boiling (nucleation of vapor due to temperature changes) is another very similar nucleation phenomenon.
We note that we use a broad definition of cavitation (destruction of a state of negative pressure by the formation of a macroscopic bubble) that thus naturally includes the case of \emph{air seeding} frequently discussed in plants, where the initial origin of the cavitation bubble is the aspiration of a meniscus through a membrane or porous structure.
In fact, the cavitation dynamics phenomena that we describe in this chapter are mostly independent on the microscopic mechanism that leads to nucleation.

\subsection*{Negative pressure and cavitation in plants}

\begin{figure}[htbp]
  \begin{center}
  \includegraphics[scale=0.62]{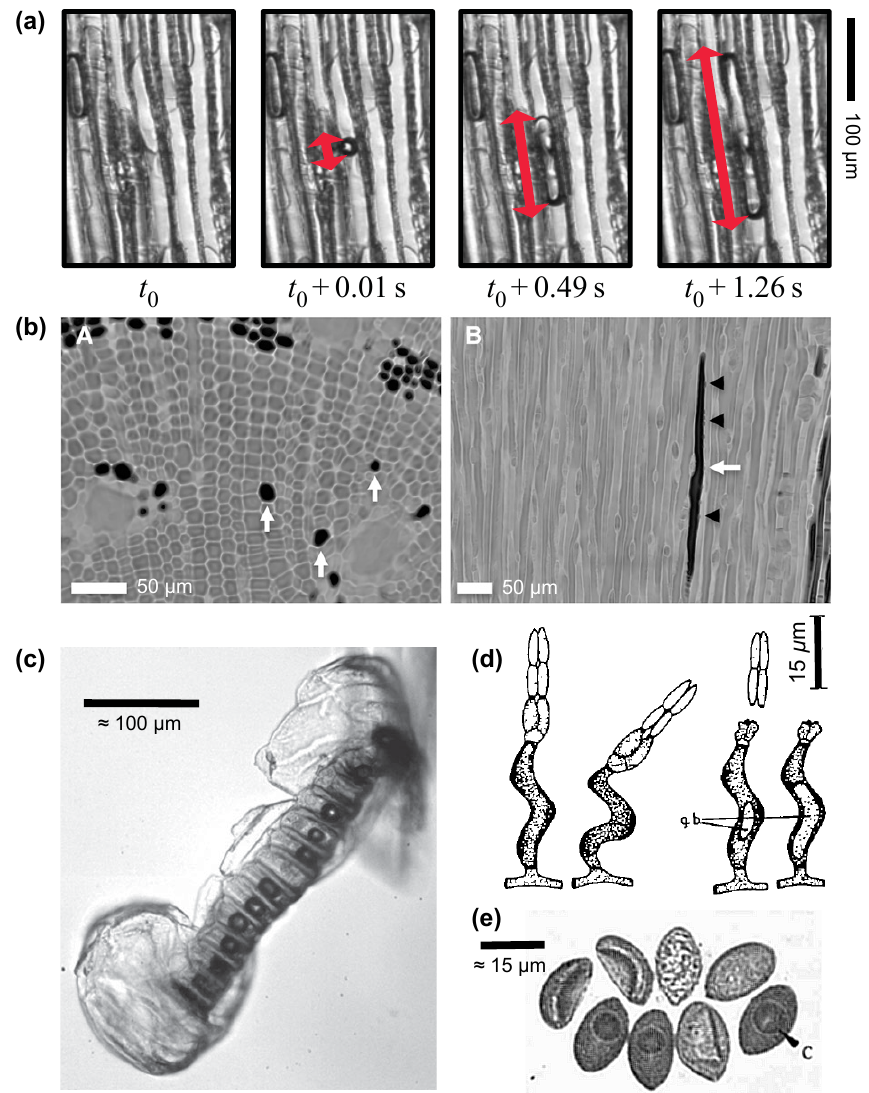}
  \caption{\small
  (a) Birth and growth of a cavitation bubble in a dehydrating slice of vascular tissue (xylem) of a pine tree.
  (b) In-vivo observation of the resulting embolism using X-ray microtomography (transverse and longitudinal cross-sections on left and right, respectively); black areas are gas-filled.
  (c) Simultaneous cavitation in neighboring cells of a fern sporangium, at the basis of a catapult-like spore ejection mechanism.
  (d) Another cavitation-based spore ejection mechanism in a fungus.
  (e) Cavitation induced by osmotic dehydration in fungal spores.
  Panels (a) and (c) republished with permission of the Royal Society, from \cite{Ponomarenko2014} and \cite{Llorens2016} respectively; permission conveyed through Copyright Clearance Center, Inc. Panels (b) and (d) reproduced from \cite{Choat2016} and \cite{Meredith1963} respectively, by permission of Oxford University Press. Panel (e) reproduced from \cite{Milburn1970} by permission of Wiley, \copyright New Phytologist Trust.
  }
\label{fig:CavitationInPlants}
\end{center}
\end{figure}

In the context of plants, water at negative pressure is ubiquitous.
Vascular plants like trees actually use that negative pressure as a suction force to drive sap flow from the root to the leaves in the \emph{xylem} tissue.
This process is driven by transpiration, i.e. natural evaporation from the xylem into the atmosphere in the leaves \cite{Tyree2013, Cochard2006, Stroock2014}.
Most trees routinely experience tens of atmospheres of negative pressure in their xylem during the day \cite{Cochard2006, Dumais2012, Stroock2014} and thus contain large amounts of water under stress, which is metastable and susceptible to cavitation (see Figure \ref{fig:CavitationInPlants}a).
Here, cavitation is detrimental, because it eventually results in air-filling (\emph{embolism}) of xylem conduits, which disrupts the upwards flow of sap (see Figure \ref{fig:CavitationInPlants}b).
Spreading of embolism through the whole vascular structure would be catastrophic, but xylem presents a segmented structure made of a large number of conducting elements (\emph{vessels}, \emph{tracheids}) interconnected by nanoporous membranes (\emph{pits}); if one of the elements fails by cavitation and gets embolized, spreading to the adjacent elements is prevented by the pits (at least temporarily), and embolism is contained (see e.g. Figure \ref{fig:CavitationInPlants}b, right panel where a single embolized tracheid is surrounded by intact water-filled tracheids).

In other situations, however, plants use cavitation to their advantage.
For example, some species of ferns use simultaneous cavitation in an annular structure as a trigger for a catapult-like ejection mechanism for their spores (\cite{Noblin2012, Llorens2016}, see Figure \ref{fig:CavitationInPlants}c).
There, the slow build-up of negative pressure prior to cavitation also originates from natural dehydration of the cells in the annulus due to evaporation into the surrounding air when conditions get dry.
Not far from the plant kingdom, negative pressure and cavitation are also observed in fungi (see Figure \ref{fig:CavitationInPlants}d-e), where they can be at the basis of other spore dispersion strategies (\cite{Money2009, Milburn1970}, see Figure \ref{fig:CavitationInPlants}d).

We briefly note that negative pressure and cavitation can also be found in the animal kingdom.
In particular, several species of shrimp use cavitation bubbles as a hunting tool (pistol shrimp \cite{Versluis2000}, mantis shrimp \cite{Patek2004}); octopuses and squids can also generate a few atmospheres of negative pressure and cavitation is thought to limit their adhesion strength \cite{Kier2002}.

The situations described above for plants and fungi have common features. First, water under negative pressure is contained in closed, cellular structures that are rigid enough to sustain the inner tension force without collapsing. Second, these structures allow for exchange of water with the surrounding environment (other cells or air) through the porous walls and/or through specialized porous membranes (pits). Last, negative pressure is generated by dehydration (transpiration at the leaf level for trees, evaporation through cell walls for the other structures) and sustains typical values in the range $-10$ to $-100$ bars for large amounts of time.

\subsection*{Chapter contents}

In this chapter, we discuss the general physics of negative pressure and cavitation in water for situations similar to those found in plant and fungi: large, static negative pressures produced by dehydration in a liquid confined in porous, cellular structures.

First, we introduce important concepts related to the mechanics and thermodynamics of water (section \ref{sec:WaterProperties}).
Then, we explain how negative pressures are generated, focusing on the effect of dehydration from cells (section \ref{sec:NegativePressureOrigin}).
Next, we discuss various microscopic mechanisms that may lead to cavitation in liquids at negative pressure (section \ref{sec:CavitationMechanisms}).
In a fourth section, we present an extension of classical nucleation theory that describes cavitation in closed, elastic cells (section \ref{sec:NucleationTheory}).
The last two sections illustrate the various theoretical concepts established previously with experimental results on artificial systems reproducing some features found in plants: we first describe the rich dynamics of a cavitation in a single cell, that spans several orders of magnitude of timescales (section \ref{sec:BubbleDynamics}), before discussing the spatio-temporal patterns of cavitation (and its propagation) in extended systems containing many, interconnected cells (section \ref{sec:CavitationPropagation}).

This chapter presents a physicist's perspective on the basics of negative pressure and cavitation in plant-like, cellular structures and thus focuses on a local level.
There are many other aspects of water relations and flow in plants that are not discussed here (e.g. global sap flow, stomatal control in the leaves, exchange between xylem and phloem, etc.).
For more details on the physiology and water transport mechanisms in plants and fungi, we refer the reader to classic books \cite{Tyree2013, Nobel2020, Money2009} and recent reviews on various aspect of water stress and transport in plants \cite{Cochard2006, Dumais2012, Stroock2014, Jensen2016}. We also recommend exhaustive books \cite{Debenedetti1996, Brennen2014} and reviews \cite{Caupin2006, Caupin2013} on the physics of metastable water and cavitation.

\section{Water properties \label{sec:WaterProperties}}

In this section, we introduce and define several properties of water that are relevant in the context of negative pressure and for the phenomenon of cavitation.

\subsection{Cohesion and surface tension\label{sec:CohestionSurfaceTension}}

Liquid water has large internal cohesion, which is mostly due to the presence of hydrogen bonds.
As a result, since molecules strongly bond together, exposing water molecules at an interface is highly unfavorable, and the surface tension of liquid water is particularly high:
$\gamma = 72 \, \mathrm{mN / m}$
at 25{\degree}C.
As a comparison, the surface tension of organic liquids such as ethanol or acetone is less than a third of that value ($\sim 22$ mN/m).

From the value of the surface tension, one can illustrate liquid water's cohesion with the following thought experiment.
Imagine "pulling" on water sufficiently strongly to separate the molecules further than their typical interaction distance, $\delta$, thus "breaking open" a liquid column.
The energy associated with the newly created interfaces, $E = 2 \gamma S$, must be equal to the work $W \simeq F_\mathrm{pull} \times \delta$ of the force pulling the liquid apart.
One can thus estimate that in order to "break open" a water column it takes a stress $\Delta P_\mathrm{pull} = F_\mathrm{pull} / S = 2 \gamma / \delta \sim 300$ MPa, assuming $\delta \sim 0.5$ nm.
From this rough estimate, we can already see that liquid water should be able to withstand very large tensile stresses, i.e. negative pressure. We refine this estimate in section \ref{sec:Compressibility}.

Note that contrary to liquids, gases do not have internal cohesion.
As a result, they cannot be brought to a state of negative pressure: "pulling" on a gas will just dilute it towards the limit of zero pressure for infinite expansion.
A direct consequence of this remark is that a liquid at negative pressure cannot coexist with a gas, except in situations where something (the surface tension of an interface, a membrane, etc.) allows for a pressure mismatch between the two phases.

\subsection{Compressibility and spinodal\label{sec:Compressibility}}

Liquid water under negative pressure is stretched, and the amount by which it stretches is related to the compressibility of the liquid.
Thus, compressibility plays a natural role in cavitation phenomena and will be involved in several of the sections of this chapter.
Also, as we discuss briefly at the end of this section, the evolution of compressibility with pressure provides a way to estimate the tensile strength of the liquid through its \emph{spinodal}.

The value of the isothermal compressibility, $\chi_\ell = (-1/V_\ell)(\partial V_\ell / \partial P)_T$, for liquid water is
$\chi_\ell = 0.45 \, \mathrm{GPa}^{-1}$ i.e. its bulk modulus is $K_\ell = 1/\chi_\ell = 2.2$ GPa; in other words, a pressure of $22$ MPa will contract water's volume, $V_\ell$, by $1\%$.

Generally, as long as the variations of pressure are not too large ($| \chi_\ell \Delta P | \ll 1$), it is a good approximation to relate the variation of volume of the liquid to variations of pressure through the linear approximation
\begin{equation}
  \left( V_\ell - V_{\ell,\mathrm{ref}} \right) / V_{\ell,\mathrm{ref}}  = - \chi_\ell \left( P - P_\mathrm{ref} \right)
\label{eq:Compressibility}
\end{equation}
where we consider deviations from a reference state at volume and pressure $(V_{\ell,\mathrm{ref}}, P_\mathrm{ref})$.
In other words, Equation (\ref{eq:Compressibility}) assumes $\chi_\ell$ constant in the range of pressure of interest.
This linear approximation expressed by Equation (\ref{eq:Compressibility}) is largely sufficient to describe the situations we consider in this chapter, where the negative pressures have a magnitude of at most $\simeq 20$ MPa, i.e. two orders of magnitude below $K_\ell = 1 / \chi_\ell$.

For negative pressures of larger magnitude, it would be necessary to take into account the variations of $\chi_\ell$ with $P$.
In fact, it is interesting to note that the variation of $\chi_\ell$ with pressure offers another way to estimate the cohesive tensile strength of liquid water.
Indeed, going back to our thought experiment from above (section \ref{sec:CohestionSurfaceTension}), compressibility is the reason why molecules get further apart when we try to "pull" on a liquid.
However, because the range of interaction of the molecules is not infinite, there must be a point where the interaction starts getting weaker and cannot resist the pulling force anymore.
At this point, tensile stress as a function of volume reaches a maximum, in other words the bulk modulus becomes zero and the compressibility $\chi_\ell$ diverges.
A way to estimate the tensile strength of the liquid is thus to extrapolate the equation of state of the liquid to estimate where $K_\ell = 1 / \chi_\ell$ is zero.
Following this approach, values of tensile strength in the vicinity of $200$ MPa (i.e. a negative pressure of $-200$ MPa) can be found at ambient temperature \cite{Speedy1982, Caupin2013}.
This validates the rough estimate made above from the value of the liquid's surface tension.

The value of $P \simeq -200$ MPa actually corresponds to the \emph{spinodal} of the liquid, where it becomes mechanically unstable and spontaneously breaks apart.
In practice, this point is never reached because cavitation appears beforehand by various mechanisms (see section \ref{sec:CavitationMechanisms}).

\subsection{Saturation pressure, phase diagram}

\begin{figure}[]
  \begin{center}
  \includegraphics[scale=0.85]{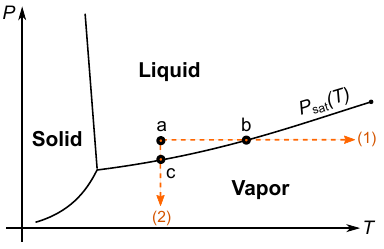}
  \caption{\small
  Schematic Pressure-Temperature phase diagram of water.
  $P_\mathrm{sat}(T)$ is the saturation vapor pressure, which represents the coexistence line between the zones of stability of the liquid and vapor phases.
  When a liquid (point a) is heated up (path 1), vapor can become the stable phase (beyond point b): vapor bubbles can spontaneously nucleate, corresponding to \emph{boiling}.
  When pressure is decreased from point a (path 2), vapor can also become the stable phase (below point c) and the nucleation of vapor bubbles is called \emph{cavitation} in this case.
  }
\label{fig:PTDiagram}
\end{center}
\end{figure}

Above we have discussed mechanical properties of water. Now, we move to thermodynamic considerations.
Generally, the most stable phase of water depends on conditions of pressure and temperature.
It is useful to represent these conditions as a point in the pressure-temperature ($P$-$T$) phase diagram (see Figure \ref{fig:PTDiagram}).
The diagram defines three zones corresponding to the zones where ice, liquid water and water vapor are the most stable, respectively\footnote{In reality, more zones are present in the ice domain because there exists different types of ice.}.
The line separating the zones of stability of the vapor and liquid phases between the triple point and the critical point defines the \emph{saturation vapor pressure}, $P_\mathrm{sat}(T)$; in ambient conditions (e.g. $T=25{\degree}$C), $P_\mathrm{sat} \simeq 3$ kPa, a value much smaller than the atmospheric pressure, $P_\mathrm{atm} \simeq 100$ kPa.

Most of us are familiar (e.g. when cooking pasta) with the fact that when heated up, water boils at $100{\degree}$C because one crosses the liquid-vapor coexistence line and one enters the domain where the vapor is most stable (see Figure \ref{fig:PTDiagram}, path 1).
Boiling occurs because for $T>100{\degree}$C, $P_\mathrm{sat}(T)$ is larger than atmospheric pressure $P_\mathrm{atm}$.
The coexistence line can also be crossed by decreasing the pressure (see Figure \ref{fig:PTDiagram}, path 2).
In this case, we do not call the nucleation of vapor bubbles \emph{boiling}, but \emph{cavitation}.
Cavitation occurs because the liquid pressure falls below the value of $P_\mathrm{sat}$ at ambient temperature.
In practice, both boiling and cavitation do not necessarily happen exactly when the coexistence line $P_\mathrm{sat}(T)$ is crossed, because of the existence of metastable states.

\subsection{Metastable states}

The liquid to vapor phase transition requires the formation of a bubble, i.e. the creation of a liquid-vapor interface.
Due to the large surface tension of water, this process is particularly unfavorable and an energy barrier (and a critical bubble size, see Introduction) must be overcome to form a bubble.
As a result, the liquid state can persist even when the coexistence line is crossed (i.e. for $P < P_\mathrm{sat}$).
This situation is metastable: small fluctuations relax back to the liquid state, but a large, less likely fluctuation can make the system cross the energy barrier and create a stable bubble, i.e. induce cavitation or boiling.
The further away from the coexistence line one penetrates into the zone of vapor stability in the P-T phase diagram, the smaller the energy barrier is and the more likely nucleation is; see the section on nucleation theory for an illustration of these aspects (section \ref{sec:NucleationTheory}).

Similar metastable states exist for the vapor to liquid phase transition (supersaturated vapor), or the liquid to ice transition (supercooled liquid)\footnote{For example, supercooled water can be brought down to $\simeq -40{\degree}C$ before nucleation of ice occurs. Supercooled water plays a large role in the atmosphere and for the survival of some living organisms in the winter \cite{Debenedetti1996}}.
Metastable liquid water below $P_\mathrm{sat}$ is often called \emph{superheated} when following the boiling route, and \emph{stretched} when following the cavitation route.
Indeed, lowering the pressure results in an expansion of the liquid due to water's compressibility (see section \ref{sec:Compressibility}).

\subsection{Chemical potential, water potential \label{sec:ChemicalWaterPotential}}

Thermodynamic driving forces and equilibria are conveniently described by the \emph{chemical potential} of water, $\mu$ [J/mol], which combines energetic and entropic contributions and enables comparison of the relative stability of the same substance in different phases: liquid water ($\mu_\ell$), water vapor ($\mu_\mathrm{v}$).
Molecules spontaneously move from areas of high $\mu$ to areas of low $\mu$ and equilibrium is attained when the chemical potential is identical in all parts of a system.
In particular, at the coexistence line ($P = P_\mathrm{sat}(T)$), $\mu_\ell(P_\mathrm{sat}) = \mu_\mathrm{v}(P_\mathrm{sat}) = \mu_\mathrm{sat}(T)$.

The concept of \emph{water potential} ($\Psi$, in Pa) is more widely used in the plant science community instead of chemical potential:
\begin{equation}
  \Psi = \frac{\mu - \mu_\mathrm{ref}}{v_\mathrm{m}}
  \label{eq:Psi}
\end{equation}
where $\mu_\mathrm{ref}(T)$ is the chemical potential of a reference state usually chosen to be pure liquid water at atmospheric pressure ($P = P_\mathrm{atm}$), at the temperature of interest, $T$; $v_\mathrm{m}$ is the molar volume of liquid water in the reference state ($v_\mathrm{m} = 1.807 \times 10^{-5} \, \mathrm{m^3/mol}$ at $T = 25 \degree C$).
Thus, chemical potential and water potential only differ by a multiplicative factor and an additive constant, and fundamentally represent the same quantity. Water potential is convenient to use, because it has units of pressure and is zero for bulk water at ambient conditions.

The reason of using $P = P_\mathrm{atm}$ instead of $P = P_\mathrm{sat}$ as a reference state is not often discussed, but this choice has physical meaning. Indeed, $P_\mathrm{sat}$ corresponds to the equilibrium between liquid water and pure water vapor; in this situation, both phases are at the same pressure. Now, let us consider liquid water situated in air at atmospheric pressure (e.g. in an open bottle) and evaluate equilibrium with water vapor in the air. The liquid phase, due to mechanical equilibrium with air, is at $P = P_\mathrm{atm}$, while the vapor phase is characterized by its \emph{partial vapor pressure}, $p \neq P$. Due to this pressure mismatch, equilibrium between the two phases occurs at a different vapor pressure, $p_\mathrm{sat} \neq P_\mathrm{sat}$ in the presence of air: the condition for equilibrium is $(p = p_\mathrm{sat}, P = P_\mathrm{atm})$, with $p_\mathrm{sat} > P_\mathrm{sat}$ (see Appendix \ref{sec:AppendixAirVaporPressure}). The difference between $p_\mathrm{sat}$ and $P_\mathrm{sat}$ is small and difficult to measure in practice (e.g. at $25 \degree C$, $P_\mathrm{sat}=3170$ Pa, $p_\mathrm{sat}=3172$ Pa, i.e. a difference of $0.07\%$), so that both saturation pressures are usually assumed to be the same. However, it is useful to keep the distinction between the two definitions in mind, in order to avoid thermodynamic inconsistencies in calculations.

The conclusion from the paragraph above is that while $P_\mathrm{sat}$ corresponds to liquid-vapor equilibrium of the pure water substance (i.e. in vacuum), the equilibrium state with liquid water at total pressure, $P=P_\mathrm{atm}$ and with water vapor at partial pressure, $p=p_\mathrm{sat} > P_\mathrm{sat}$ is a true saturation state of the system when considering water in air at atmospheric pressure. It is thus natural to use this state as a reference when dealing with processes occurring in air.
Using this reference, Equation (\ref{eq:Psi}) yields
\begin{equation}
  \Psi_\ell = \Delta P - \Pi
  \label{eq:PsiLiq}
\end{equation}
for liquid water, where $\Delta P = P - P_\mathrm{atm}$ is the pressure difference with respect to atmospheric pressure, and $\Pi$ is the \emph{osmotic pressure}; $\Pi = 0$ for pure water, and $\Pi \simeq RTC > 0$ when solutes are present at a concentration $C \, \mathrm{[ mol/m^3]}$ \cite{Nobel2020,Stroock2014,Vincent2019}.
$\Delta P$ is a purely mechanical contribution, while $\Pi$ is an entropic contribution associated with the colligative decrease of chemical potential that occurs when mixing water with a solute.
Here, we neglect the contribution of gravity ($\Delta \Psi_\mathrm{g} = \rho g z$, with $\rho$ the density of liquid water, $g$ the acceleration of gravity and $z$ the elevation) that will not be needed for the discussions in this chapter, and is only significant to compare water at large height differences, e.g. in tall trees ($\rho g \simeq 0.01 \, \mathrm{MPa / m}$). We have also neglected the contribution due to the compressibility of water. Indeed, for consistency with Equation (\ref{eq:Compressibility}), an additional term $\Delta \Psi_\mathrm{\chi} = \chi_\ell \Delta P^2 / 2$ should be included. However this contribution is also negligible for the conditions considered in this chapter ($ | \Delta \Psi_\chi / \Psi | < 0.5 \, \%$ for $\Psi > -20$ MPa), we thus neglect this term for simplicity.

For water vapor in air,
\begin{equation}
  \Psi_\mathrm{v} = \frac{RT}{v_\mathrm{m}} \ln \left( \frac{p}{p_\mathrm{sat}} \right)
  \label{eq:PsiVap}
\end{equation}
where $p_\mathrm{sat}$ is the saturation vapor pressure \emph{in air} defined above; $a = p / p_\mathrm{sat}$ is the \emph{activity} of water vapor or its \emph{relative humidity}: for example $a = 0.85$ corresponds to a relative humidity of $85 \%$RH.
Although this is rarely mentioned, $\Psi_\mathrm{v}$ is impacted by gravity in the exact same way as $\Psi_\ell$ (i.e., $\Delta \Psi = \rho g z$, where $\rho$ is still the density of liquid water)\footnote{The fact that the density of the liquid, $\rho$, is involved in the equation for water vapor is simply due to the way water potential is defined, i.e. by dividing water potential differences $\Delta \mu$ by the molar volume of the \emph{liquid}; for water vapor, $\Delta \mu = RT \ln (a) + M g z$, where $M$ is the molar mass of water. Not taking into account the effect of gravity on the water potential of water vapor can result in the erroneous prediction that one can build a perpetual-motion pumping machine based on liquid/vapor local equilibrium, i.e. a tall column of liquid water could never be in equilibrium with the surrounding vapor.}.
Again, we neglect this effect here.

Similarly to chemical potential ($\mu$), water potential ($\Psi$) is the natural driving force for transport and equilibria within a single phase or between phases. Transport occurs towards areas of lower $\Psi$ and equilibrium occurs when the driving force vanishes, i.e. equal $\Psi$ in different parts of the system.

\subsection{Evaporation vs. cavitation \label{sec:EvaporationVsCavitation}}

Equation (\ref{eq:PsiVap}) implies that the water potential of water vapor in air is always negative when relative humidity is less than 100\%RH ($a<1$). This means that an open container with pure, liquid water ($\Psi_\ell = 0$, see equation \ref{eq:PsiLiq}) always tends to evaporate, except in rare cases of 100\%RH humidity, because water molecules lower their water potential by moving from the liquid phase to the vapor phase. This corresponds to the fact that when $a<1$, there are more molecules per unit time leaving the liquid towards the vapor than molecules coming back from the vapor into the liquid, so that over time the liquid loses mass.
Evaporation thus occurs at the interface between liquid and air due to a transport imbalance between the liquid and vapor phases. In the example above, the two phases (liquid at $P_\mathrm{atm}$, vapor at $p <= p_\mathrm{sat}$) are both in their stability zone of the phase diagram, but evaporation can also occur from a metastable liquid into a subsaturated vapor through a membrane (see section \ref{sec:DehydrationThermo}).
Contrary to evaporation, cavitation and boiling correspond to the nucleation of the stable vapor phase \emph{within} the metastable liquid, and do not happen in a stable liquid.

\section{Origins of negative pressure \label{sec:NegativePressureOrigin}}

As explained in the introduction, negative pressure in plants mainly originates from dehydration (e.g. evaporation, osmosis).
Below, we first explain the connection between negative pressure and dehydration from a mechanical perspective.
Then, we describe the driving forces and equilibrium states involved in dehydration from a thermodynamic perspective.
Finally, we briefly mention other ways negative pressures can be obtained.

\subsection{Dehydration: mechanics \label{sec:DehydrationMechanics}}

Let's consider liquid water that is enclosed in a cell of volume, $V_\mathrm{c}$, at ambient pressure $P_\mathrm{atm}$.
Now, we examine a situation where the cell dehydrates, i.e. loses water molecules (by an amount of substance $\Delta n$), e.g. by transport through the porous walls. We assume that liquid adhesion to the cell walls is good so that no cavitation occurs and the liquid remains homogeneous in the process.

\begin{figure}[]
  \begin{center}
  \includegraphics[scale=0.72]{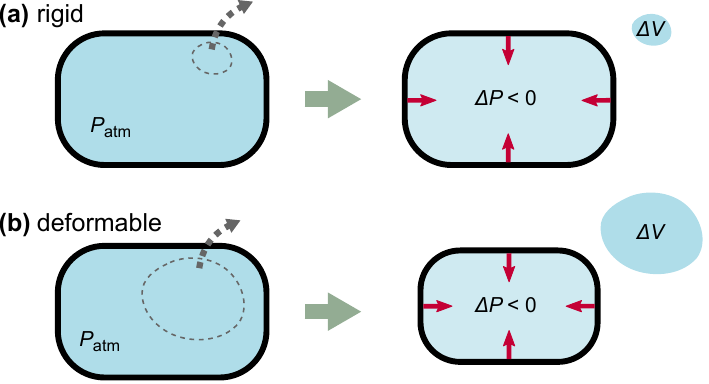}
  \caption{\small
  Mechanics of negative pressure generation by dehydration of a cell containing liquid water.
  (a) If the confining cell is infinitely rigid, withdrawing a volume $\Delta V$ of the liquid stretches the liquid by the same amount, which makes the pressure drop (see Equation \ref{eq:DehydrationMechanicsRigid}).
  (b) If the cell is deformable, the same process both stretches the liquid and makes the cell shrink. More liquid volume needs to be removed to produce a similar drop in pressure (see Equation \ref{eq:DehydrationMechanicsGeneral}).
  Red arrows represent the force exerted by the liquid on the walls;
  lighter blue color indicates a less dense liquid.
  }
\label{fig:DehydrationMechanics}
\end{center}
\end{figure}

If the cell is infinitely rigid (see Figure \ref{fig:DehydrationMechanics}a), $V_\mathrm{c}$ does not change in the process, and the cell now contains less molecules occupying the same volume.
Necessarily, the average distance between molecules in the liquid increases; in other words the liquid gets stretched and its pressure drops: $P = P_\mathrm{atm} + \Delta P$, with $\Delta P < 0$.
Another way to look at the situation is that the molecules initially present in the cell (volume, $V_\mathrm{c}$) now occupy a larger volume, $V_\mathrm{c} + \Delta V$, where $\Delta V = v_\mathrm{m} \Delta n$ is the liquid volume occupied by the molecules removed from the cell.
We can evaluate the pressure required to stretch the liquid by an amount $\Delta V$ using Equation (\ref{eq:Compressibility}):
\begin{equation}
    \Delta P = - \frac{1}{\chi_\ell} \frac{\Delta V}{V}.
    \label{eq:DehydrationMechanicsRigid}
\end{equation}
Because of the low compressibility of water ($\chi_\ell = 0.45 \, \mathrm{GPa}^{-1}$, see section \ref{sec:Compressibility}), removing only a small fraction of the water molecules within the cell results in massive pressure changes.
For example withdrawing only 0.1\% of the molecules ($\Delta V / V = 0.001$) is sufficient to make the pressure drop from ambient to more than 20 bars of negative pressure, $\Delta P = -2.2$ MPa.

In real systems, the cell enclosing the liquid is not infinitely rigid and deforms under liquid pressure changes.
In particular, if $P$ decreases, the cell tends to contract so that $V_\mathrm{c}$ also decreases (see Figure \ref{fig:DehydrationMechanics}b).
We can quantify this effect using an effective cell compressibility $\chi_\mathrm{c} = (1/V_\mathrm{c})(\partial V_\mathrm{c} / \partial P)_T$; note the opposite sign in this definition compared to the liquid's compressibility (see section \ref{sec:Compressibility})) that allows us to keep all coefficients positive.
$K_\mathrm{c} = 1 / \chi_\mathrm{c}$ is the effective bulk modulus of the cell.
Now, removing a volume $\Delta V = v_\mathrm{m} \Delta n$ of liquid water from within the cell results in both contraction of the cell and stretching of the liquid. Using the definitions of $\chi_\ell$ and $\chi_\mathrm{c}$, it is straightforward to show that the corresponding drop in pressure is given by
\begin{equation}
    \Delta P = - \frac{1}{\chi_\ell + \chi_\mathrm{c}} \frac{\Delta V}{V}
    \label{eq:DehydrationMechanicsGeneral}
\end{equation}
as long as the pressure variations are small with respect to both bulk moduli ($| \chi_\ell \Delta P | \ll 1$, $| \chi_\mathrm{c} \Delta P | \ll 1$). We recall that $\Delta V$ is the liquid volume withdrawn from the cell, not the change in the cell volume, $\Delta V_\mathrm{c}$; $\Delta V$ and $\Delta V_\mathrm{c}$ are different in general because of the compressibility of the liquid.
Because $\chi_\mathrm{c}$ is positive, it requires more dehydration of the cell (larger $\Delta V$) to achieve a similar drop in pressure compared to the case of an infinitely rigid cell discussed previously, as can be seen by comparing eqns \eqref{eq:DehydrationMechanicsRigid} and \eqref{eq:DehydrationMechanicsGeneral}.

Equation (\ref{eq:DehydrationMechanicsGeneral}) shows the interplay of water compressibility ($\chi_\ell$) and cell deformability ($\chi_\mathrm{c}$).
When dehydration occurs, water expansion and cell contraction occur simultaneously.
One can have different views on the generation of negative pressure by dehydration.
Some people would prefer to consider that water leaving the cell induces an elastic strain in the cell wall, which "pulls" on the liquid and generates negative pressure.
Other people would think of the same process as water leaving the cell that stretches the inner liquid, resulting in negative pressure that pulls on the cell walls and make the cell shrink.
However, these are just two point of views on a unique, coupled phenomenon.
Note that even with a very stiff cell, a tiny deformation of the walls has to occur to maintain equilibrium with the pressure force exerted by the liquid.

We have assumed a linear response between the pressure and the cell volume, which is an excellent approximation as long as $|\Delta P | \ll K_\mathrm{c}$ (small deformations).
For larger deformations, one would need to know in more detail the stress-strain relationship of the cell to quantify deviations to Equation (\ref{eq:DehydrationMechanicsGeneral}).
Also, instabilities such as creasing or buckling can occur (some are visible in Figure \ref{fig:CavitationInPlants}e) and may result in collapse of the structure.
To avoid too large deformations (that reduce the available volume) or collapse, structures conducting water at negative pressure should be stiff ($K_\mathrm{c} \gg |\Delta P |$).
In trees (see Figure \ref{fig:CavitationInPlants}a-b), for example, typical negative pressures in xylem are $-1$ to $-10$ MPa, while $K_\mathrm{c}$ is in the range $0.1 \-- 1$ GPa; deformations of wood under the effect of water negative pressure are thus small but measurable \cite{Louf2017,Irvine1997}.
For cavitation-based spore ejection mechanisms (see Figure \ref{fig:CavitationInPlants}c-d), however, the structure is much more compliant, which allows the system to deform significantly and store a large amount of elastic energy when negative pressure builds up; this stored energy is suddenly released when cavitation occurs. In these situations, the cell elastic moduli and the negative pressures are of the same order of magnitude ($\sim 10$ MPa)\footnote{For the fern sporangium \cite{Llorens2016}, typical negative pressures of $-10$ MPa develop, and the cell elastic modulus is $K_\mathrm{c} = 2B / h$ where $B \simeq 400$ N/m is the bending modulus of the annulus, and $h \simeq 40 \, \mathrm{\mu m}$ is the height of the cells. Thus, $K_\mathrm{c} \simeq 20$ MPa.}.

\subsection{Dehydration: thermodynamics \label{sec:DehydrationThermo}}

\begin{figure}[]
  \begin{center}
  \includegraphics[scale=0.72]{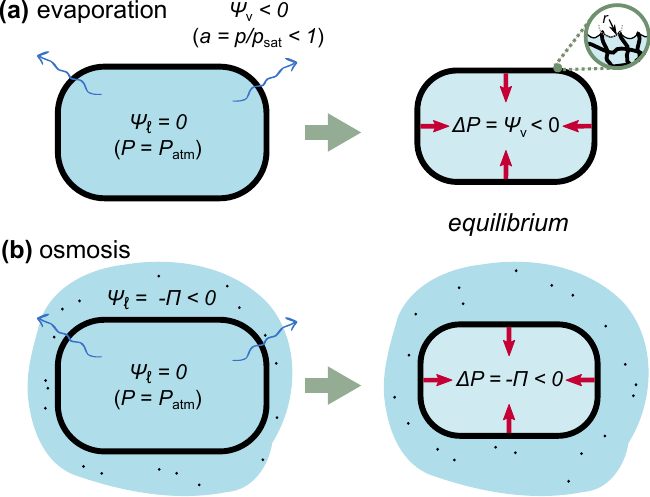}
  \caption{\small
  Thermodynamics of negative pressure originating from dehydration.
  (a) Evaporation in a subsaturated vapor dehydrates the cell until the negative pressure in the cell balances the negative water potential of the external water vapor (see Equation \ref{eq:Kelvin}).
  Equilibrium between the liquid and vapor at different pressures can be mediated through the formation of curved liquid-vapor menisci (inset).
  (b) Dehydration in an osmotic solution produces the same effect; equilibrium is reached when the negative pressure in the cell balances the external osmotic pressure (see text).
  }
\label{fig:DehydrationThermo}
\end{center}
\end{figure}

In the discussion above, we have explained how dehydration can generate negative pressure through the coupled stretching of the liquid and strain of the enclosing structure that it implies. Now, we discuss the thermodynamic driving forces for dehydration and the potential associated equilibria.

As discussed in section \ref{sec:ChemicalWaterPotential}, driving forces can be described using \emph{water potential}, $\Psi$.
In the same way as a drop of water ($\Psi_\ell = 0$) that spontaneously evaporates in a subsaturated vapor ($a = p / p_\mathrm{sat} < 1$, hence $\Psi_\mathrm{v} < 0$) as discussed in section \ref{sec:EvaporationVsCavitation}, a cell structure containing water, initially at ambient pressure ($\Psi_\ell = 0$ if there are no solutes), tends to dehydrate spontaneously if the membrane permits exchange with air that is subsaturated in water vapor (see Figure \ref{fig:DehydrationThermo}a).
From the mechanical mechanisms described in the previous section (see Figure \ref{fig:DehydrationMechanics}), this dehydration produces a drop in pressure, which decreases the water potential of the liquid within the cell (see Equation \ref{eq:PsiLiq}).
As dehydration progresses, $\Psi_\ell$ within the cell becomes more and more negative (as does the liquid pressure) and gets closer to $\Psi_\mathrm{v}$ of the vapor, so that the driving force for dehydration decreases.
An equilibrium can be reached when $\Psi_\ell = \Psi_\mathrm{v}$, or, from Equations (\ref{eq:PsiLiq})-(\ref{eq:PsiVap}),
\begin{equation}
  \Delta P = \frac{RT}{v_\mathrm{m}} \ln \left( \frac{p}{p_\mathrm{sat}} \right) + \Pi
  \label{eq:Kelvin}
\end{equation}
where $\Delta P = P - P_\mathrm{atm}$, and $\Pi$ is the osmotic pressure associated with potential solutes present in liquid water.
In trees, xylem contains solutes at low concentrations, and the associated osmotic pressure is usually neglected \cite{Tyree2013,Nobel2020}.
However, other structures such as a the fern leptosporangium can contain solutes with $\Pi$ in the MPa range \cite{Llorens2016};
in such a situation, as can be seen in Equation (\ref{eq:Kelvin}), the presence of solutes makes the equilibrium pressure at given external humidity less negative than for pure water.

For pure water, the negative pressure implied by Equation (\ref{eq:Kelvin}) is rapidly increasing in magnitude as the relative humidity of the air is decreased below 100\%RH (see Table \ref{table:Kelvin}) so that equilibrium with a subsaturated vapor is a very efficient way to generate massive negative pressures.
With solutes, the values of liquid pressure indicated in Table \ref{table:Kelvin} have to be shifted up by the amount $\Pi$ (e.g., the equilibrium pressure of a cell containing solutes at $\Pi = 2$ MPa at $92.9 \%$RH would be $-8$ MPa).

\begin{table*}[]
  \centering
  \begin{tabular}{| c || c | c  | c | c | c | c | c |}
   \hline
   air humidity (\%RH)       & 100                     & 99.93  & 99.2  & 92.9  & 83.3 & 69.4  & 48.2 \\
   \hline
   liquid pressure (MPa)     & 0.1 ($=P_\mathrm{atm}$) & 0      & -1    & -10   & -25  & -50   & -100 \\
   \hline
   radius of curvature (nm)  & $\infty$                & 1498   & 131   & 14    & 5.7  & 2.9   & 1.4  \\
   \hline
  \end{tabular}
  \caption{\small
  Equilibrium liquid pressure, $P = P_\mathrm{atm} + \Delta P$, and radius of curvature, $r$, of the liquid/vapor interface as a function of water vapor humidity in the air (\%RH = $p / p_\mathrm{sat} \times 100$), calculated from Equations (\ref{eq:Kelvin}-\ref{eq:Laplace}) for a temperature of $T = 298.15$ K.
  }
  \label{table:Kelvin}
\end{table*}

We note that dehydration can also be driven by osmosis instead of evaporation. Taking the example of a cell containing pure liquid water immersed in a solution at ambient pressure (see Figure \ref{fig:DehydrationThermo}b), if the cell membrane is able to maintain the concentration difference (i.e. excludes the solute), equilibrium is obtained when a the negative pressure in the inner liquid balances the osmotic pressure, i.e. $\Delta P = -\Pi$. This strategy is sometimes used to artificially induce negative pressure and cavitation in plant cells or artificial structures (see e.g. \cite{Milburn1970, Llorens2016, Vincent2012a, Vincent2014, Scognamiglio2018, Vincent2019}). More generally, with solutes both inside and outside of the cell, $\Delta P = \Pi_\mathrm{in} - \Pi_\mathrm{out}$ at equilibrium.
The dehydration effect still occurs if the membrane only hinders solute transport instead of being fully permeable, but with a reduced efficiency: $\Delta P = \sigma (\Pi_\mathrm{in} - \Pi_\mathrm{out})$, with the \emph{reflection coefficient} $\sigma < 1$ \cite{Deen1987,Nobel2020,Vincent2019}.
If not actively maintained, the concentration difference however tends to vanish over time due to solute diffusion, and the osmotic pressure difference becomes zero.
In some sense, osmosis can be seen as less efficient than evaporation as a driving force to induce negative pressures, because large concentrations of solutes are required to generate significant osmotic pressures (e.g. $1$ mol/L of sucrose for $\Pi \simeq 3$ MPa \cite{Nobel2020}), while small changes in humidity generate large negative pressure differences (see Table \ref{table:Kelvin}).

Of course, the negative pressure described by Equation \ref{eq:Kelvin} is only reached if equilibrium can be attained. This means that the membrane separating the liquid at negative pressure ($P_0 + \Delta P$, $\Delta P < 0$) and the air (pressure, $P_0$) must resist collapse and/or mechanical failure under the pressure difference $\Delta P$, but also allow for the coexistence of liquid and air at very different pressures. One way to achieve the latter is if the membrane is wetted by the liquid and allows curved liquid-vapor menisci to develop (see Figure \ref{fig:DehydrationThermo}a, inset).
Assuming for simplicity that the menisci are hemispherical caps with the same curvature $1/r$ in both directions, they maintain a pressure difference
\begin{equation}
  \Delta P = - 2 \gamma / r
  \label{eq:Laplace}
\end{equation}
from Laplace law. The lower the humidity, the larger the pressure difference and the larger the curvature of the menisci.
Note that the combination of Equations (\ref{eq:Kelvin}) and (\ref{eq:Laplace}) is known as the Kelvin-Laplace equation.
Table \ref{table:Kelvin} indicates radii of curvature predicted from the Kelvin-Laplace equation; in order for the curvature to develop, the constrictions within the membrane (pore radius $r_\mathrm{p}$ for a porous medium, typical half mesh size of the polymer network for a hydrogel, etc.) must be smaller than $r$, assuming the medium to be well wetted by water.
For less hydrophilic surfaces, the constrictions must be even smaller (e.g. for a pore, $r_\mathrm{p} < r \times \cos \theta$, where $\theta$ is the receding water contact angle on the pore wall).

Of course, meniscus invasion in a complex nanoporous structure cannot be simply represented by a single constriction size (and the concept of a meniscus is somewhat debatable in polymer structures), so these estimates must be taken as order of magnitudes only.
However, the values in Table \ref{table:Kelvin} indicate that only membranes containing networks with typical dimensions in the nanometer range are suitable to be able to develop large negative pressures.
Static negative pressures described by Equation (\ref{eq:Kelvin}) down to $-20$ to $-25$ MPa could in fact be obtained in artificial systems consisting in large voids ($\sim 10 \-- 100 \, \mathrm{\mu m}$ in size) separated from air by membranes based on hydrophilic hydrogels \cite{Wheeler2008,Wheeler2009,Vincent2012a} or hydrophilic mesoporous silicon \cite{Vincent2014a, Chen2016, Vincent2019}, both with estimated constrictions below $5$ nm.

In fact, in the experiments mentioned in the previous paragraph based on hydrogels or mesoporous silicon, the mechanism limiting the achievable negative pressure to $-25$ MPa was not capillary failure in the membrane but cavitation of the containing liquid.
More generally, the equilibrium described by Equation (\ref{eq:Kelvin}) is metastable when the pressure is negative, because of the possibility of bubble nucleation by various mechanisms (see section \ref{sec:CavitationMechanisms}).
It is remarkable that as long as the system allows for equilibrium (no mechanical or capillary failure of the membrane, no cavitation), the equilibrium predicted by Equation (\ref{eq:Kelvin}) does not depends on the details of the membrane itself.

\subsection{Dehydration: discussion}

The values of negative pressure listed in Table \ref{table:Kelvin} could make one think that trees should routinely endure massive water stresses and cavitation. Indeed, it is not rare for humidity to fall below $50 \%$RH during the day, which corresponds to $\sim -100$ MPa of pressure at equilibrium, close to the homogeneous cavitation limit (see section \ref{sec:CavitationMechanisms}).
In reality, water within xylem is not at equilibrium with the atmosphere, first because the water-conducting cells are continuously fed with water coming from the roots, and also because trees have protection mechanisms that isolate water in xylem from the outside when pressure becomes too negative (e.g. closure of stomates \cite{Tyree2013, Nobel2020}).
As a result, the actual negative pressure in xylem arises from a balance between water potential in the soil, water potential in the atmosphere, and the different resistances to water transport separating the soil and the atmosphere through the xylem tissue \cite{Tyree2013, Nobel2020, Caupin2013, Jensen2016}.
However the general idea that dehydration (in this case by transpiration in the leaves) is the driving force for negative pressure generation remains true; simply the local water potential (i.e dehydration state) in the xylem is not as severe as one could expect, due to continuous flow or isolation.
Other natural structures that rely on cavitation for spore ejection (see Figure \ref{fig:CavitationInPlants}) are in more direct contact with the atmosphere and potentially quickly reach equilibrium with the local humidity surrounding them. As a result, they require the humidity to drop below a critical value for their mechanisms to trigger.

\subsection{Other origins of negative pressure}

Although negative pressure in plants originates from dehydration-related processes in natural conditions, there exist other methods to induce negative pressure and cavitation.
We only give a brief summary here, more details can be found in other references \cite{Caupin2006,Caupin2013,Vincent2012a}.

\paragraph{Traction}

This is perhaps the most intuitive way to stretch a liquid, i.e. by literally pulling on it, for example using the piston in a syringe.
Any air pocket must be carefully removed because the pulling force would just make the gas expand without pulling on the liquid.
As noted by Huygens in the 17th century, the liquid's own weight in a vertical tube can also lead to a tensile state and two centuries later Reynolds obtained negative pressures of a few atmospheres using this method \cite{Caupin2006}.

\paragraph{Hydrodynamics}

Accelerating a fluid can result in a drop in pressure (Venturi effect) sufficiently large to make the pressure negative.
This effect is well-known by mechanical engineers, because cavitation around fast-spinning boat propellers results in damage from the violent collapse of the bubbles on the blades \cite{Franc2005}.
The order of magnitude of negative pressure achievable with this technique is typically of a few bars at maximum, probably due to remaining air bubbles in the fluid that act as cavitation nuclei \cite{Morch2007}

\paragraph{Centrifugation}

Spinning a tube containing a liquid, around an axis perpendicular to the length of the tube, makes the liquid "pull on itself" due to the centrifugal effect, resulting in a gradient of pressure with the minimum value at the rotation axis.
Briggs obtained negative pressures down to $-28$ MPa with this technique, which has also been adapted to study cavitation in xylem by spinning branches \cite{Holbrook1995,Cochard2002,Cochard2013}.

\paragraph{Isochoric cooling}

Known as the Berthelot method \cite{Berthelot1850}, this technique consists in cooling down a liquid trapped in a pocket within a solid: the liquid would naturally contract, but if there is good adhesion with the walls, the liquid is forced to remain at a volume larger than at equilibrium and gets stretched.
After cavitation, the pocket of liquid can be "reset" by heating up the system sufficiently so that the liquid expands again and occupies the whole volume.
The rare experiments that reached what seems to be cavitation by homogeneous nucleation at $-120$ to $-140$ MPa used the Berthelot technique with micrometric inclusions in quartz \cite{Zheng1991,Azouzi2013}.

\paragraph{Acoustic waves}

Acoustic waves are an oscillation of pressure around ambient pressure. If the amplitude of the wave is large enough (e.g. due to strong focalization in a liquid), the low-pressure part of the oscillation cycle can be below $P=0$.
Despite the fact that the probed volume and timescales are small and that the wave can be focalized away from any surface, even careful experiments don't seem to be able to reach pressures below $-30$ MPa \cite{Herbert2006,Caupin2006}.
One experiment using reflected shock waves reports pressures down to $-60$ MPa, albeit with a more indirect way of estimating the pressure, based on comparison with simulations of shockwave propagation \cite{Ando2012}.

\section{Cavitation Mechanisms \label{sec:CavitationMechanisms}}

\begin{figure*}[]
  \begin{center}
  \includegraphics[scale=0.9]{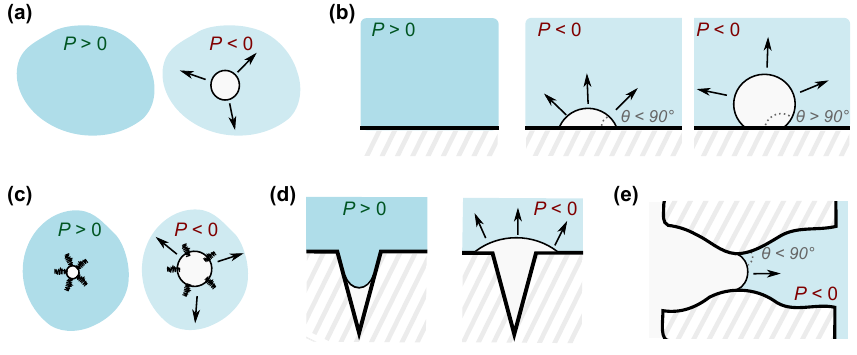}
  \caption{\small
  Examples of cavitation mechanisms.
  (a) Homogeneous nucleation.
  (b) Surface-aided nucleation (left: hydrophilic surface; right: hydrophobic surface).
  (c) Cavitation from a floating gas bubble (black elements represent potential organic molecules stabilizing the surface of the bubble).
  (d) Cavitation from air trapped in a crevice.
  (e) Cavitation from meniscus aspiration through a pore.
  Black arrows represent motion, dashed areas represent solids. Bubbles contain gas (water vapor and/or air).
  }
\label{fig:CavitationMechanisms}
\end{center}
\end{figure*}

Although water is theoretically able to withstand $\sim -200$ MPa of pressure before spontaneously breaking apart (spinodal limit, see section \ref{sec:Compressibility}), cavitation always occurs at more moderate values of negative pressure, so that the spinodal is never reached in practice.
The pressure at which cavitation happens, $P_\mathrm{cav}$ strongly depends on the microscopic mechanism leading to the formation of a bubble.
Below, we briefly introduce some commonly discussed cavitation mechanisms.

\subsection{Homogeneous nucleation}

Cavitation by homogeneous nucleation (see Figure \ref{fig:CavitationMechanisms}a) corresponds to the spontaneous nucleation of a vapor bubble in the liquid from thermal fluctuations.
Although this process is obviously stochastic, the probability of cavitation is very nonlinear as a function of pressure and varies very quickly around a typical value of $P_\mathrm{cav}$.
Theoretical estimates for $P_\mathrm{cav}$ are usually based on classical nucleation theory (CNT) \cite{Blander1975,Debenedetti1996} and are typically in the range $-120$ to $-150$ MPa at ambient temperatures, depending on considered volumes, timescales and refinements of CNT \cite{Caupin2006,Azouzi2013,Menzl2016}.
In section \ref{sec:Compressibility}, we present an extended version of CNT that takes into account finite volume and the effect of compressibility (of the liquid and of the containing cell).

The limit $P_\mathrm{cav} \sim -120$ to $-150$ MPa is at more moderate negative pressures than the spinodal ($\sim -150$ to $-200$ MPa \cite{Speedy1982, Caupin2013}) and can be seen as the ultimate negative pressure that one can hope to reach in water, as even thermal fluctuations are sufficient to produce cavitation at that point.
Reaching this limit experimentally is difficult, as one must eliminate all other mechanisms that can trigger nucleation at less negative pressures (see examples below).
The rare experiments that have reached such extreme negative pressures have used thermal cycling in water pockets trapped in quartz inclusions \cite{Zheng1991,Azouzi2013}.
Many other methods (including acoustic waves focused in a pure liquid far from any walls) seem to find a cavitation threshold at $-20$ to $-30$ MPa, even with careful purification and preparation; there is no clear understanding on the reason of such a difference between these experiments and those with quartz inclusions \cite{Caupin2013}.

The other mechanisms discussed below can be all considered as \emph{heterogeneous} mechanisms, by opposition to the homogeneous nucleation mechanism described here.

\subsection{Surface-aided nucleation}

Homogeneous nucleation is difficult because creating a liquid-vapor interface costs energy.
Any solid surface that the liquid does not completely wet (i.e. contact angle, $\theta > 0$) reduces this energetic cost by enabling pathways of nucleation where the bubble is a truncated sphere (see Figure \ref{fig:CavitationMechanisms}b).
This effect can easily be incorporated into CNT \cite{Blander1975, Caupin2006}; the predicted $P_\mathrm{cav}$ depends on $\theta$ but only deviates significantly from that of homogeneous nucleation when the surface is hydrophobic ($\theta > 90 \degree$).
For example, a contact angle greater than $150 \degree$ would be required to predict $P_\mathrm{cav} = -20$ MPa \cite{Vincent2012a}.
Globally, any cavitation pressure can be obtained between the homogeneous nucleation value ($-120$ to $-150$ MPa) and $P = P_\mathrm{sat}$ by varying the contact angle between $0$ and $180 \degree$. Non-flat surfaces (e.g. crevices or tips) can also modify the cavitation pressure \cite{Wilt1986}.

Fundamentally, the surface-aided mechanism is not very different from that of homogeneous nucleation in the sense that it still involves a vapor bubble that is spontaneously nucleated in the liquid, but with a reduction in energy cost due to a surface.
Cavitation through this mechanism can be seen as a \emph{loss of adhesion} between the liquids and the surface.
The other mechanisms discussed below, however, are fundamentally different because they involve a pre-existing germ; the liquid-vapor interface already exists and is simply expanded when cavitation occurs.

\subsection{Seeded cavitation}

Here we discuss several cavitation mechanisms that involve pre-existing germs, i.e. microscopic bubbles that can contain gas (typically, air) or water vapor, which are stable at moderate negative pressures, but become unstable at some critical pressure value $P_\mathrm{cav}$ where they expand into a macroscopic bubble, i.e. cavitation.

For a gas bubble in a liquid, $P_\mathrm{cav} = P_\mathrm{sat} - 4\gamma / 3 R$, where $R$ is the radius of the bubble\footnote{Obviously, the radius of the bubble depends on the pressure in the surrounding liquid. Here, the radius to consider is the one when $P = P_\mathrm{cav}$; solving for $P_\mathrm{cav}$ requires using the equilibrium condition given by Laplace law \cite{Brennen2014, Vincent2017a}.}, which is known as the \emph{Blake threshold}.
The cavitation pressure of a liquid containing bubbles is thus dictated by the largest bubble it contains.

It is worth noting, however, that a bubble containing gas that is soluble in the surrounding liquid is never stable and spontaneously dissolves due to the combined effect of surface tension and diffusion \cite{Epstein1950,Brennen2014}.
Air bubbles dissolve quickly (e.g. $\sim 1$ minute for an initial radius of $\sim 10 \, \mathrm{\mu m}$) so in the context of plants where negative pressures last for hours, it is difficult to imagine situations where air bubbles would trigger cavitation before having completely dissolved. Gas germs thus have to be stabilized against dissolution in some way.

One way is for the bubble to have an organic coating (see Figure \ref{fig:CavitationMechanisms}c) that blocks diffusion and/or reduce surface tension. Blake's threshold is supposed to still be a good estimate of $P_\mathrm{cav}$ in this situation \cite{Atchley1989, Morch2007}.
Another stabilizing mechanism is that of air trapped in a crevice (see Figure \ref{fig:CavitationMechanisms}d): with some conditions on geometry and surface wettability, such a bubble can be stable, and gets "extracted" from the crevice if the pressure gets too low in the liquid, resulting in cavitation \cite{Harvey1944,Apfel1970, Trevena1987,Wheeler2009}.
Even with a conical crevice, there are many cases to consider, but the result is that, again, any cavitation pressure between $P_\mathrm{sat}$ and the homogeneous nucleation threshold can be obtained by varying the crevice size, angle, gas content and wettability.

A related mechanism is that of a pore separating the liquid at negative pressure from an air-filled (or vapor-filled) space (see Figure \ref{fig:CavitationMechanisms}e).
The situation becomes unstable if the pressure difference $\Delta P = P - P_\mathrm{gas}$ falls below a critical value $\Delta P_\mathrm{cav}$, which is set by the maximum magnitude of the Laplace pressure that the geometry can allow: for a cylindrical pore of radius $r_\mathrm{p}$, $\Delta P_\mathrm{cav} = - 2 \gamma \cos \theta / r_\mathrm{p}$\footnote{For a surface that exhibits contact angle hysteresis, $\theta$ would represent the receding contact angle in this situation.}.
This simple estimate can be adapted for more realistic pore geometries, e.g. for pit membranes in xylem, with the interesting conclusion that $\Delta P_\mathrm{cav}$ not only depends on the pore size distribution but also strongly on the thickness of the membrane \cite{Kaack2021}.
Nevertheless, the simple equation for a cylindrical pore allows us to illustrate the general idea that cavitation occurs at more negative pressures for smaller and more wettable pores.
This mechanism is often referred to as \emph{air seeding} in the plant science literature.

Because $\Delta P$ can also be lowered by increasing $P_\mathrm{gas}$ without changing the liquid pressure $P$, air seeding can be triggered by pressurizing the gas phase instead of putting the liquid under tension.
In fact, external air pressurization is one of the popular methods used by plant scientists to estimate vulnerability of xylem to cavitation, although it can lead to various artifacts \cite{Cochard2013}.
One could argue that air seeding is not a \emph{true} cavitation mechanism, because it does not require the liquid to be thermodynamically metastable (e.g. in the situation of air pressurization described above).
However, in natural situations the gas pressure is not artificially changed, and the pore sizes (e.g. in xylem pits) are sufficiently small (sub-micron) to allow for liquid pressures lower than $P_\mathrm{sat}$ (i.e. thermodynamically metastable) to develop in the liquid.
In such situations, the formation of a bubble by air seeding falls under our definition of cavitation, as the expansion of the bubble makes the liquid relax from a metastable state to a stable state.

\subsection{Discussion}

From the description of cavitation mechanisms above, we can extract the following conclusions:
\begin{itemize}
  \item The most negative pressure achievable in water at ambient temperature is set by homogeneous nucleation and is larger than one thousand atmospheres ($P_\mathrm{cav}^\mathrm{(hom)} = -120$ to $-150$ MPa) in magnitude.
  \item $-20$ to $-30$ MPa seems to be a more practical lower limit, while plants typically develop only $-1$ to $-10$ MPa.
  \item Heterogeneous mechanisms can set the cavitation pressure of a system to any value between $P_\mathrm{sat}$ and $P_\mathrm{cav}^\mathrm{(hom)}$
  \item Resistance to cavitation is better if defects in the system (pores or crevices in the walls, hydrophobic patches, floating bubbles, etc.) are smaller (and/or more wettable when applicable).
\end{itemize}
This last point is illustrated by the fact that cavitation pressure scales as $\gamma / L$, where $L$ is a typical size (e.g. bubble radius for Blake's threshold, pore size for air seeding, critical bubble radius for homogeneous nucleation, see section \ref{sec:NucleationTheory}).

It is usually difficult to pinpoint a specific cavitation mechanism, but some mechanisms can be ruled out by various tests.
For example, cavitation seeded by air trapped in crevices can be modified or even suppressed by pre-pressurization of the system.
Examples can be found of such tests in artificial, xylem-like systems based on hydrogels \cite{Wheeler2009} or porous silicon \cite{Chen2016}, with a careful discussion of potential cavitation mechanisms.

Other cavitation mechanisms may exist. For example, recent results show that a negative pressure of $\simeq -10$ MPa is sufficient to tear apart lipid bilayers, which could thus be responsible for cavitation by loss of adhesion within the bilayer due to the tensile force of water acting on it \cite{Kanduc2020}. More generally, there has been a growing interest in understanding the role of surfactants present in sap or in cell walls on cavitation in plants \cite{Schenk2015,Schenk2017}.

Generally, the size separation between the microscopic germ responsible for cavitation and the final cavitation bubble is so large that the exact cavitation mechanism has little impact on the dynamics of the cavitation bubble after nucleation.
As a result, the dynamical phenomena (bubble dynamics, cavitation propagation, etc.) presented in the end of this chapter are mostly insensitive to the microscopic mechanism at the origin of the cavitation bubble.

\section{Confined Cavitation Theory \label{sec:NucleationTheory}}

\begin{figure*}[]
  \begin{center}
  \includegraphics[scale=0.72]{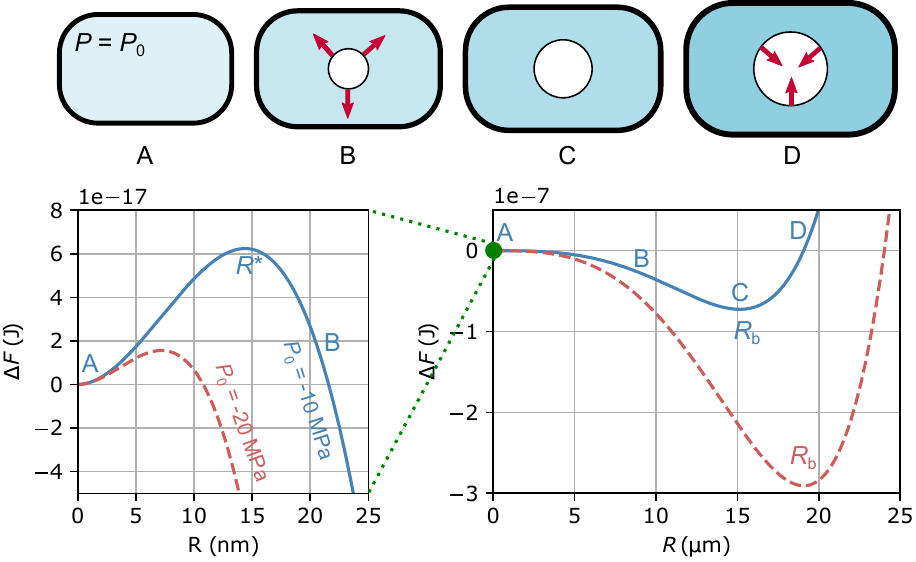}
  \caption{\small
  Cavitation in a closed container (cell), assuming a spherical bubble and an effective container compressibility of $\chi_\mathrm{c} = 1 \, \mathrm{GPa}^{-1}$.
  State A corresponds to the homogeneous liquid at negative pressure ($P_0 < 0$).
  In state B, a bubble has overcome the critical nucleation radius $R^\ast$: the effective pressure $P_\mathrm{eff}$, representing the net driving force applied on the bubble (red arrows), is negative and tends to make the bubble grow.
  State C corresponds to equilibrium ($P_\mathrm{eff} = 0$, $R = R_\mathrm{b}$).
  In state C, the bubble is larger than its equilibrium size and a net inwards force ($P_\mathrm{eff}$) tends to make it go back to its equilibrium size.
  Oscillations between states B and D around equilibrium are possible due to inertia in the system.
  In the schematics, darker shades of blue indicate larger pressure and larger density in the liquid.
  In the graphs, the continuous, blue line corresponds to an initial negative pressure of $P_0 =-10$ MPa while the dashed, red curve corresponds to $P_0 = -20$ MPa.
  The two graphs represent the same $\Delta F (R)$ function (Equation \ref{eq:FreeEnergyNucleation}) on different ranges of bubble size. Notice the large scale separation between them.
  }
\label{fig:ConfinedNucleation}
\end{center}
\end{figure*}

Here, we present a single theoretical framework that will allow us to discuss both thermodynamic aspects and dynamical aspects of negative pressure and cavitation in closed environments.
Mainly, it consists in an extension of classical nucleation theory (CNT) \cite{Blander1975,Debenedetti1996,Caupin2006,Brennen2014} to take into account the compressibility of the liquid and the elasticity of the containing cell \cite{Vincent2012a,Vincent2017a}.
As we have shown in section \ref{sec:NegativePressureOrigin}, these two effects are important in the generation of negative pressure. We will show below that they also play a significant role for the equilibrium states and dynamics of cavitation bubbles.

The idea is to evaluate the minimum work of formation of a bubble in a liquid at negative pressure in a closed, elastic container (see Figure \ref{fig:ConfinedNucleation}).
The minimum work identifies with the \emph{Helmholtz free energy} difference, $\Delta F$ between an initial state without bubble and with the liquid at negative pressure, $P_0 < 0$ (see Figure \ref{fig:ConfinedNucleation}, state A) and a a state with a bubble of volume, $V$ and surface area, $A$ (states B-D).
$\Delta F$ also corresponds to the potential energy of the system and dictates the driving forces that shape the dynamics.

We consider a bubble forming within a liquid, i.e. homogeneous nucleation (see section \ref{sec:CavitationMechanisms} for a discussion of other mechanisms). We also assume that the liquid does not contain dissolved air, or that if it does, transport of air by diffusion into the bubble is negligible during the timescales of cavitation. As a result, the saturation vapor pressure of interest is $P_ \mathrm{sat}$ (but using $p_\mathrm{sat}$ would result in negligible differences anyway, see section \ref{sec:ChemicalWaterPotential}). It can be shown (see Appendix \ref{sec:AppendixNucleationTheory} for a detailed derivation) that in this situation,
\begin{equation}
  \Delta F = \gamma A + (P_0 - P_\mathrm{sat}) V + \frac{1}{2} \frac{1}{V_0 (\chi_\ell + \chi_\mathrm{c})} V^2
  \label{eq:FreeEnergyNucleation}
\end{equation}
where $V_0$ is the initial container volume (equivalently, initial liquid volume), $\chi_\ell$ is the isothermal compressibility of liquid water introduced previously, and $\chi_\mathrm{c}$ the effective compressibility of the container (see section \ref{sec:DehydrationMechanics}).

In Equation (\ref{eq:FreeEnergyNucleation}), the first term represents the energetic cost of creating a liquid-vapor interface (surface tension, $\gamma$), the second term is associated with the initial driving force associated with the pressure in the liquid, and the third term describes how this driving force changes as a function of bubble size due to compressibility effects. Indeed, Equation (\ref{eq:FreeEnergyNucleation}) assumes that during the timescales of interest ($\sim \, \mathrm{\mu s}$, see below), no significant transport happens through the walls of the containing cell, so that the number of water molecules in the cell remains constant. As a result, the only way to make the bubble grow is by compressing the surrounding liquid and/or inflating the containing cell, hence the $\chi_\ell + \chi_\mathrm{c}$ factor in the expression.
The compression of the liquid as the bubble grows results in an increase in pressure in the liquid (see Equation \ref{eq:VolumesPressureVolumeBubble}): the negative pressure progressively vanishes as a result.
The assumption of short timescales also means that the value of $\chi_\mathrm{c}$ to consider might be different than in quasi-static situations, e.g. for poroelastic systems \cite{Wang2017a}. Generally, $\chi_\mathrm{c}$ represents here the pressure/volume relation of the containing cell at the current conditions of dehydration in the medium, and in a no-flow situation (constant water content in the cell).

Equation (\ref{eq:FreeEnergyNucleation}) is applicable for any bubble shape. In the following, we assume the bubble to be spherical, so that $A = 4 \pi R^2$ and $V = 4 \pi R^3 / 3$, where $R$ is the bubble radius. The quantity
\begin{equation}
    P_\mathrm{eff} = \frac{\mathrm{d} \Delta F}{\mathrm{d} V} = \frac{2 \gamma}{R} + (P_0 - P_\mathrm{sat}) + \frac{1}{\chi_\ell + \chi_\mathrm{c}} \frac{V}{V_0}
    \label{eq:EffectivePressureNucleation}
\end{equation}
is an effective pressure that represents the driving force for the bubble dynamics. If $P_\mathrm{eff} < 0$, the system is in tension and the bubble tends to expand (Figure \ref{fig:ConfinedNucleation}, state B). Equilibrium occurs when $P_\mathrm{eff} = 0$ (Figure \ref{fig:ConfinedNucleation}, state C). If the bubble is larger than equilibrium (Figure \ref{fig:ConfinedNucleation}, state D), the driving force reverses and tends to make the bubble shrink back to equilibrium.

Figure \ref{fig:ConfinedNucleation} presents the shape of $\Delta F$ for two different initial negative pressures, using typical values $\chi_\mathrm{c} = 1 \, \mathrm{GPa}^{-1}$ and $V_0 = (100 \, \mathrm{\mu m}^3)$ for the containing cell, and tabulated values for the properties of water at $25 \degree C$ ($\gamma = 0.072$ N/m, $P_\mathrm{sat} = 3170$ Pa, $\chi_\ell = 0.45 \, \mathrm{GPa}^{-1}$).
Below, we discuss the main features of the  $\Delta F$ potential energy landscape and their consequences.

\subsection{Critical radius, energy barrier}

For small radii in the nanometer range, $\Delta F$ presents a maximum at the \emph{critical radius}, $R^\ast$ (see Figure \ref{fig:ConfinedNucleation}, left).
Below $R^\ast$, $P_\mathrm{eff} > 0$ and the bubble tends to collapse.
Only when $R > R^\ast$ does the negative pressure "win" over surface tension ($P_\mathrm{eff} < 0$), making the bubble expand and resulting in cavitation.
Thus, $\Delta F^\ast = \Delta F(R^\ast)$ corresponds to the \emph{energy barrier} for nucleation. This energy barrier decreases when pressure becomes more negative, making nucleation more likely.

With typical container sizes (dimensions $0.01 \-- 1$ mm and above) for plant cells, the compressibility contribution (third terms in Equations \ref{eq:FreeEnergyNucleation}-\ref{eq:EffectivePressureNucleation}) can be shown to be negligible around the critical radius, so that $\Delta F$ is well approximated by the first two terms, which correspond to those usually considered in CNT for homogeneous nucleation \cite{Caupin2006}.
In this situation, the critical radius and energy barrier are easily calculated from Equations (\ref{eq:FreeEnergyNucleation}-\ref{eq:EffectivePressureNucleation}), neglecting the last term and requiring $P_\mathrm{eff} = 0$; this calculation yields  $R^\ast = 2 \gamma / (P_\mathrm{sat} - P_0)$ and $\Delta F^\ast = 4 \pi {R^\ast}^2 \gamma / 3$.

In sub-micron container sizes, however, the surface tension and compressibility terms can become of the same order of magnitude around $R^\ast$, with the surprising consequence that a liquid at negative pressure confined at these dimensions can become absolutely stable instead of metastable \cite{Vincent2012a,Vincent2017a}.

The classical approach of CNT is to compare the energy barrier $\Delta F^\ast$ to thermal fluctuations ($k_\mathrm{b} T$) and compute the probability of passing the energy barrier given some attempt frequency. Doing so, one estimates that the probability of cavitation is virtually zero unless pressure reaches values in the $-120$ to $-150$ MPa range \cite{Caupin2006,Azouzi2013,Menzl2016}.
These values depend only weakly on total volume and observation timescales.

The expression (\ref{eq:FreeEnergyNucleation}) of $\Delta F$ can be modified to describe cavitation seeded by gas germs, by adding a term $- 3 N k_\mathrm{b} T \ln V$; in this case, nucleation occurs from a gas bubble containing $N$ insoluble gas molecules and a confined version of Blake's threshold (see section \ref{sec:CavitationMechanisms}) can be obtained \cite{Vincent2017a}.
More generally, the shape of $\Delta F$, the values of the critical radius and energy barrier and the associated typical cavitation pressure depend greatly on the microscopic mechanism of cavitation (see section \ref{sec:CavitationMechanisms}). The shape of the potential $\Delta F(R)$ for radii much larger than the critical radius, however, does not depend significantly on the cavitation mechanism. In the rest of this chapter, we focus on this large-bubble regime, which describes the formation of a macroscopic cavitation bubble once the energy barrier is overcome.

\subsection{Equilibrium bubble \label{sec:EquilibriumBubble}}

At large bubble sizes, $\Delta F$ presents a minimum at a point corresponding to the equilibrium bubble (see Figure \ref{fig:ConfinedNucleation}, state C). We define the equilibrium bubble volume as $V_\mathrm{b}$ and corresponding radius as $R_\mathrm{b}$. Once the energy barrier of nucleation is overcome, the expanding cavitation bubble will tend towards $V_\mathrm{b}$.
The equilibrium size of the bubble is found by solving $\mathrm{d}\Delta F / \mathrm{dV} = 0$, i.e. $P_\mathrm{eff} = 0$. Using Equation (\ref{eq:EffectivePressureNucleation}),
\begin{equation}
    V_\mathrm{b} = V_0 \left(\chi_\ell + \chi_\mathrm{c} \right) \left( P_\mathrm{sat} - P_0 - \frac{2 \gamma}{R_\mathrm{b}} \right).
    \label{eq:EquilibriumVolumeExact}
\end{equation}
In most situations relevant to this chapter (real and artificial plant-like systems), $P_\mathrm{sat}$ and the capillary pressure $2 \gamma / R_\mathrm{b}$ are negligible in magnitude compared to $P_0$, so that
\begin{equation}
  V_\mathrm{b} \simeq V_0 \left(\chi_\ell + \chi_\mathrm{c} \right)  | P_0 |
  \label{eq:EquilibriumVolumeApprox}
\end{equation}
to excellent approximation. Equations (\ref{eq:EquilibriumVolumeExact}-\ref{eq:EquilibriumVolumeApprox}) express mathematically the fact that the volume occupied by the bubble comes from the space liberated by the combined contraction of the liquid and expansion of the cell, due to a change of pressure from $P_0 < 0$ to $P \simeq 0$. Larger bubble sizes are obtained if the initial negative pressure is larger in magnitude ($P_0$ more negative), if the cell is more deformable (higher $\chi_\mathrm{c}$), or if the cell's volume, $V_0$, is larger.

We note that with the typical parameters that we have used for the confining cell's volume and elasticity, the size of the equilibrium bubble ($R_\mathrm{b}$) is several orders of magnitude larger than the size of the critical bubble ($R^\ast$), as can be seen in Figure \ref{fig:ConfinedNucleation}.
Consequently, the critical radius and energy barrier are not visible on a graph of $\Delta F$ at the length scales and energy scales of the equilibrium bubble (see Figure \ref{fig:ConfinedNucleation}, right).
This illustrates the fact that the microscopic mechanism for nucleation is irrelevant for the dynamics of formation of the macroscopic bubble, as discussed earlier.

\subsection{Inertial oscillations \label{sec:BubbleDynamicsOscillations}}

Around the equilibrium bubble, $\Delta F$ has the shape of a potential well.
When a cavitation bubble appears, inertia can make the system oscillate around the equilibrium position at the bottom of the well (see Figure \ref{fig:ConfinedNucleation}, right, arrows).
This results in radial oscillations of the bubble.

\paragraph{Effective stiffness}

In order to evaluate the oscillation dynamics, it is useful to approximate the system as a harmonic oscillator around its equilibrium point, i.e. use a quadratic expansion of $\Delta F (R)$ around $R=R_\mathrm{b}$: $\Delta F (R) \simeq \Delta F (R_\mathrm{b}) + k (R - R_\mathrm{b})^2 / 2$, where $k = (\mathrm{d}^2 \Delta F / \mathrm{d} R^2)_{R=R_\mathrm{b}}$ is the \emph{effective stiffness} of the system. Using the definitions $P_\mathrm{eff} = \mathrm{d} \Delta F / \mathrm{d}V$ and $A = \mathrm{d} V / \mathrm{d} R$, combined with the equilibrium condition $P_\mathrm{eff}(R_\mathrm{b})=0$, $k$ can be expressed as
\begin{equation}
  k = A_\mathrm{b} \left( \frac{\mathrm{d}P_\mathrm{eff}}{\mathrm{d}R} \right)_{R=R_\mathrm{b}}
\end{equation}
where $A_\mathrm{b}$ is the bubble surface area at equilibrium.
Using Equation (\ref{eq:EffectivePressureNucleation}),
\begin{equation}
  k = \frac{A_\mathrm{b}^2}{V_0 (\chi_\ell + \chi_\mathrm{c})} - 8 \pi \gamma
  \label{eq:EffectiveStiffness}
\end{equation}

\paragraph{Effective mass}

The oscillation dynamics is set by the interplay of the system's stiffness and inertia. In general, inertia (i.e., the kinetic energy, $E_\mathrm{k}$) is complicated to evaluate because it depends on the details of the velocity field in the liquid and in the walls of the containing cell. However, the velocity field typically decreases rapidly away from the bubble surface, so that the dominant contribution to the kinetic energy usually comes from the fluid displaced close to the bubble.
As a first approximation, one can thus assume that the kinetic energy is that of a bubble oscillating in an unbounded fluid: $E_\mathrm{k} = (1/2) m (\mathrm{d}R / \mathrm{d}t)^2$, where $m = 4 \pi R_\mathrm{b}^3 \rho$ is the effective mass of the system \cite{Brennen2014}. We recall that $\rho$ is the density of the liquid. During the oscillation, the liquid is successively compressed and stretched, so that its density changes, but these variations are small and can be ignored as a first approximation. The main correction to the effective mass comes from the effect of the actual velocity field in the liquid compared to an unbounded situation, and to the potential contribution of the other parts of the system (e.g. cell walls and outside medium) to the kinetic energy. These different effects can be bundled into a corrective factor $\phi$ so that
\begin{equation}
    m = 4 \pi R_\mathrm{b}^3 \rho \phi
    \label{eq:EffectiveMass}
\end{equation}
where $\phi$ is a dimensionless factor describing deviations to the unbounded case ($\phi = 1$ for a bubble in an infinite liquid).
In the ideal situation where the cell containing the liquid is a spherical void in a solid with an effective compressibility, $\chi_\mathrm{c}$ and density, $\rho_\mathrm{s}$ that are comparable to those of the liquid ($\chi_\ell$, $\rho$) and the bubble is at the center of the cell, $\phi \simeq 1$. However, for an infinitely stiff cell that imposes a strict zero-velocity boundary condition at the cell wall, the effective mass is significantly reduced compared to Equation \ref{eq:EffectiveMass}, e.g. $\phi \simeq 0.65$ for a bubble of radius only one fifth of the container's radius \cite{Vincent2017a}.

\paragraph{Oscillation frequency \label{sec:OscillationFrequency}}

With the effective stiffness and effective mass established above, the system is analogous to a mass/spring system, and its natural angular oscillation frequency is given by $\omega = 2 \pi f = \sqrt{k/m}$.
For simplicity, we assume in the following that the surface tension contribution to the effective stiffness ($-8 \pi \gamma$) is small compared to that of the system's compressibility; this is usually a good approximation, but it has to be evaluated depending on the system's parameters \cite{Vincent2014a, Vincent2017a}. Using Equations (\ref{eq:EffectiveStiffness}-\ref{eq:EffectiveMass}),
\begin{equation}
  \omega \simeq \sqrt{\frac{4 \pi R_\mathrm{b}}{(\chi_\ell + \chi_\mathrm{c}) \rho V_0}}.
  \label{eq:OscillationFrequency1}
\end{equation}
From Equation \ref{eq:EquilibriumVolumeExact}, we also know the bubble size $V_\mathrm{b} = 4 \pi R_\mathrm{b}^3 / 3$ as a function of the initial negative pressure prior to cavitation and we can rewrite (still neglecting surface tension):
\begin{equation}
  \omega \simeq \left( \frac{3}{\rho \phi} \right)^{1/2} \left( \frac{4 \pi}{3 V_0 (\chi_\ell + \chi_\mathrm{c})} \right)^{1/3}  \left( P_\mathrm{sat} - P_0 \right)^{1/6},
  \label{eq:OscillationFrequency2}
\end{equation}
Interestingly, Equation (\ref{eq:OscillationFrequency2}) indicates that the natural oscillation frequency of a cavitation bubble depends very weakly on the initial pressure prior to cavitation (power $1/6$). The main parameters governing the oscillation frequency are the cell's volume ($V_0$), and the system's effective compressibility ($\chi = \chi_\ell + \chi_\mathrm{c}$). For the quite different geometry of a cylindrical bubble at the end of a long cylindrical tube (length $L$), it can also be shown with a similar approach that $\omega \simeq L^{-1} \sqrt{3/(\rho \chi)}$ \cite{Vincent2012a}.

Predictions similar to Equations (\ref{eq:OscillationFrequency1}-\ref{eq:OscillationFrequency2}) can be obtained from estimations of the dispersion relation of acoustic waves related to the oscillation of the bubble within the containing cell \cite{Vincent2012a,Vincent2017a,Drysdale2017,Scognamiglio2018}, and agree with experimental results \cite{Vincent2012,Vincent2014a,Vincent2017a,Scognamiglio2018} and simulations \cite{Gallo2020}. While these acoustic methods are restricted to small oscillations around equilibrium, the full nonlinear dynamics of the bubble (Rayleigh-Plesset-like equation) can also be directly calculated from the expression of $\Delta F(R)$ and of the kinetic energy, e.g. using Euler-Lagrange equations \cite{Vincent2012a,Vincent2017a}. From this Rayleigh-Plesset-like equation it is possible to show that the initial expansion velocity of the bubble when it appears is
\begin{equation}
  \left(\frac{\mathrm{d}R}{\mathrm{d}t}\right)_0 = \sqrt{\frac{2 (P_\mathrm{sat} - P_0)}{3 \rho}}
  \label{eq:BubbleExpansionVelocity}
\end{equation}
which is in fact a general result for the rate of expansion of a bubble in a liquid at pressure, $P_0$ \cite{Brennen2014}.
Equation \ref{eq:BubbleExpansionVelocity} can be used to estimate the negative pressure in the liquid prior to cavitation by measuring the initial rate of growth of the cavitation bubble \cite{Vincent2012a,Vincent2017a,Bruning2019}.

The radial vibration of the cavitation bubble emits acoustic waves in the medium, and analysis of the frequency of the acoustic signals in relation to the theoretical predictions (e.g. Equations \ref{eq:OscillationFrequency1}-\ref{eq:OscillationFrequency2}) potentially allows observers to obtain information about the local properties (cell volume, elasticity) of the system. While this idea works well in artificial structures \cite{Vincent2012a,Scognamiglio2018,Vincent2014a}, it is less straightforward to implement in a tree, because of the complexity of wave propagation in wood \cite{DeRoo2016}.

In fact, the acoustic emission related to the formation of a cavitation bubble in a liquid enclosed in a stiff container is so strong that it accounts for the majority of the dissipation of the energy initially contained in the stretched liquid (and elastically deformed container) prior to cavitation \cite{Vincent2012a,Vincent2017a}.
This effect contrasts with the case of unconfined bubbles, for which damping usually comes from viscous or thermal effects \cite{Leighton1994,Brennen2014}.
This strong dissipation results in a very quick damping of the radial vibrations, which stop after just a few oscillations, see section \ref{sec:BubbleDynamics}.

\section{Cavitation bubble dynamics \label{sec:BubbleDynamics}}

We now illustrate the theoretical concepts introduced in the previous sections with experimental results on the dynamics of cavitation bubbles appearing in a liquid at negative pressure within an enclosed space.
As we will show, this dynamics is rich, and spans many orders of magnitude of timescales, from sub-microseconds to minutes.

\begin{figure*}[]
  \begin{center}
  \includegraphics[scale=0.72]{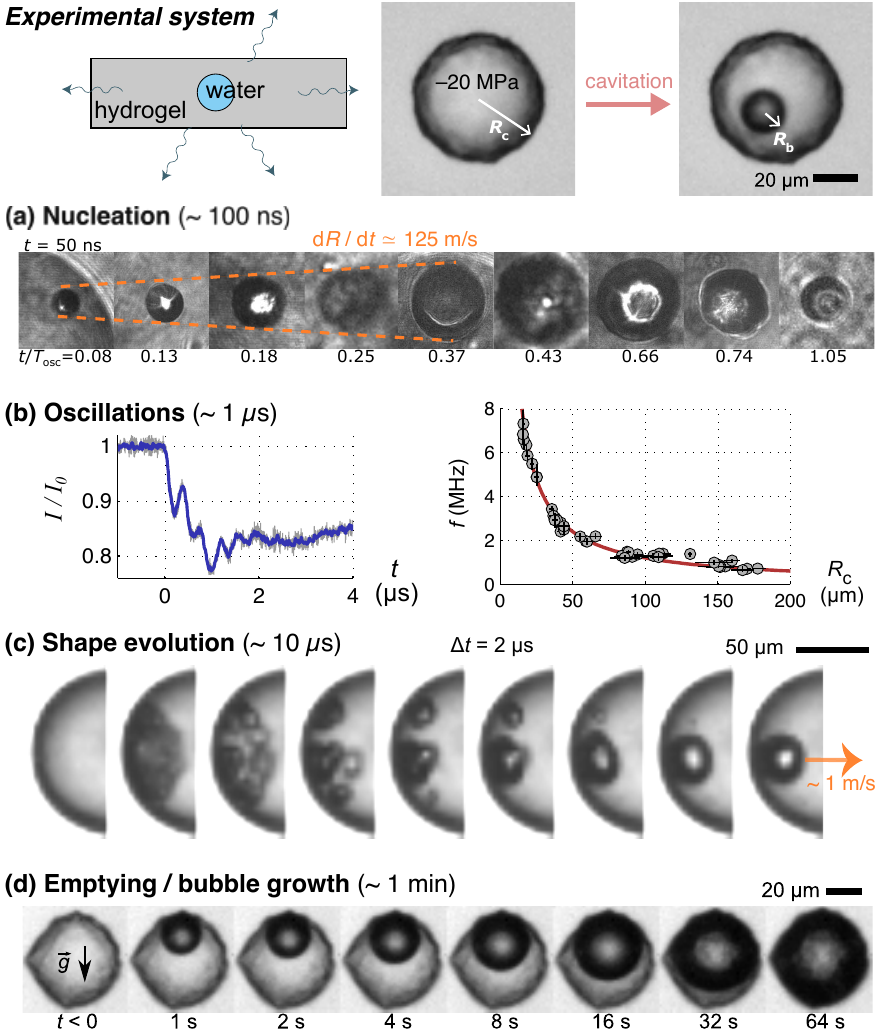}
  \caption{\small
  Cavitation bubble dynamics in an artificial cell (spherical void in a polymer hydrogel).
  (a) Nucleation dynamics reconstructed from multiple laser strobe photography experiments (time is normalized by the oscillation period, see (b)).
  (b) Inertial oscillations measured by light scattering (extinction of transmitted light), and corresponding frequency as a function of void size.
  (c) High-speed camera recordings of the evolution of the bubble shape.
  (d) Time-lapse camera recording of the final growth of the bubble and emptying of the void, from a side view.
  All panels adapted from reference \cite{Vincent2014a}, except (d) from \cite{Vincent2012a}.
  }
\label{fig:BubbleDynamics}
\end{center}
\end{figure*}

The discussion in this section is based on a series of experiments using artificial systems that mimic some essential features of cavitation in plants \cite{Vincent2012a,Vincent2012,Vincent2014a}.
The samples (see Figure \ref{fig:BubbleDynamics}, \emph{experimental system}) can be seen as an artificial cells, which consists of spherical voids filled with water, enclosed in a hydrogel (pHEMA).
Dehydration of the cells by diffusion of water through the hydrogel results in negative pressure (see section \ref{sec:DehydrationMechanics}), and is driven by evaporation (see section \ref{sec:DehydrationThermo}).
Cavitation occurs spontaneously in the liquid when $P \simeq -20$ MPa; this value is estimated by equilibrating the system with air at different relative humidities and by using Equation (\ref{eq:Kelvin}) to estimate the corresponding pressure in the liquid \cite{Wheeler2008,Wheeler2009,Vincent2012a}.
Cavitation can also be triggered at slightly higher pressures by using a laser pulse (for strobe photography measurements, see below).
The hydrogel is stiff enough to resist collapse upon the development of negative pressure in the liquid: its effective compressibility is $\chi_\mathrm{c} = 1 \, \mathrm{GPa}^{-1}$, comparable to that of xylem (see discussion in section \ref{sec:DehydrationMechanics}).
The initial development of such \emph{synthetic xylem} systems was made by Wheeler and Stroock \cite{Wheeler2008,Wheeler2009}; in particular, we refer the reader to their optimization of the mechanical properties of the gel to avoid collapse \cite{Wheeler2008} and to their thorough discussion of possible cavitation mechanisms with respect to those described in section \ref{sec:CavitationMechanisms} here \cite{Wheeler2009}.

Other experiments similar to those presented below have been reported in the literature, for example with cubic cells and osmosis-driven dehydration \cite{Scognamiglio2018}, or with spherical cells in a much softer material, PDMS ($\chi_\mathrm{c} \sim 1 \, \mathrm{MPa}^{-1}$) \cite{Bruning2019}.
In this latter case, the cell shrinks dramatically during the development of negative pressure, however the nonlinear elasticity of PDMS and the fact that cavitation occurs at a negative pressure of much lower magnitude ($P_\mathrm{cav} \sim -1$ MPa) allows the system to avoid complete collapse.

Below, we discuss the different stages of formation of a cavitation bubble in the spherical, hydrogel-based artificial cells (see Figure \ref{fig:BubbleDynamics}).

\subsection{Nucleation}

Laser strobe photography experiments consists in triggering nucleation in the metastable liquid by a first laser pulse, then illuminating the system with a second pulse separated from the first one by 50 to 500 ns.
Repeating the process several times enables an image reconstruction of the birth of the cavitation bubble (see Figure \ref{fig:BubbleDynamics}a), revealing an ultra-fast expansion.
Using Equation (\ref{eq:BubbleExpansionVelocity}), one can show that the rate of expansion of the bubble, $\mathrm{d}R / \mathrm{d}t = 125 \pm 20$ m/s \cite{Vincent2012a} is consistent with the cavitation pressure of $-20$ MPa estimated previously from humidity equilibration experiments.

\subsection{Oscillations}

After the initial expansion, the bubble starts to shrink again (see Figure \ref{fig:BubbleDynamics}a, last 3 frames), because the bubble has overshot its equilibrium position due to inertia (see section \ref{sec:NucleationTheory}): this corresponds to the onset of oscillations.
Light extinction experiments reveal oscillations with frequencies $f = \omega / (2 \pi)$ in the MHz range (see Figure \ref{fig:BubbleDynamics}b, left). This signal corresponds to radial oscillations of the bubble, that are quickly damped due to acoustic radiation (see section \ref{sec:BubbleDynamicsOscillations}).
The frequency decreases as the size of the cell is increased (see Figure \ref{fig:BubbleDynamics}b, right), in excellent agreement with theoretical predictions (Equation \ref{eq:OscillationFrequency2}, red line in Figure \ref{fig:BubbleDynamics}b, right). This frequency also depends on the compressibility of water and the elasticity of the cell, but only weakly on the cavitation pressure (see section \ref{sec:BubbleDynamicsOscillations}). Because of acoustic radiation, these oscillations can also be detected with a high-frequency hydrophone \cite{Vincent2012a,Ponomarenko2014,Scognamiglio2018}.

\subsection{Shape evolution}

A high-speed camera at half a million frames per second cannot resolve the oscillations, which result in image blur (see Figure \ref{fig:BubbleDynamics}c, second frame), but reveals a rich, slower shape-related dynamics that follows the oscillations.
The images suggest the evolution of the bubble into a toroidal shape (see Figure \ref{fig:BubbleDynamics}c, third frame) that destabilizes into several bubbles, which collapse back into a single bubble eventually in a form of quick ripening process (see Figure \ref{fig:BubbleDynamics}c, frames 4 to 9).
This peculiar dynamics is thought to arise from a combination of high-speed jets (creating the toroidal bubble \cite{Lauterborn2010}) and a Rayleigh-Plateau instability (fragmenting the toroidal bubble \cite{Vincent2014a}), but is not fully understood.
Eventually, the bubble returns to a spherical shape, while quickly moving towards the center of the cell (orange arrow); the magnitude and scaling (with bubble size) of this translation dynamics is consistent with a capillary effect driven by surface tension \cite{Vincent2012a,Vincent2014a}.

\subsection{Temporary equilibrium}

After the fast dynamics described above, the bubble relaxes to its equilibrium shape and size (see Figure \ref{fig:BubbleDynamics}, top right micrograph). The equilibrium bubble volume is proportional to the cell volume (see Equation \ref{eq:EquilibriumVolumeExact}), and is set by the elastic relaxation of both the liquid and the container (see section \ref{sec:EquilibriumBubble}).

It is worth noting that the whole process of formation of the equilibrium bubble has a duration on the order of $10 \, \mathrm{\mu s}$ and thus requires techniques with very large frame rates to capture it. A regular camera (or the naked eye) would just observe a spherical bubble suddenly appearing, floating in the bulk of the liquid, making the observer think of a homogeneous nucleation process. However, high-speed recordings such as those of Figure \ref{fig:BubbleDynamics}c strongly suggest a heterogeneous process starting on the solid walls \cite{Vincent2012a,Vincent2014a}.

The equilibrium described by the theory in section \ref{sec:NucleationTheory} assumes that the number of water molecules in the containing cell remains constant. However, the walls (here, the hydrogel) are permeable to water, and eventually an outwards flow occurs, because the driving force for dehydration (evaporation or osmosis) is still present. In other words, there is a water potential imbalance between water inside the cell, which has relaxed back to $\Psi \simeq 0$ after cavitation, and the surrounding medium ($\Psi < 0$). The initial equilibrium at volume $V_\mathrm{b}$ is thus only temporary.

\subsection{Emptying, bubble growth \label{sec:CellEmptying}}

After cavitation, the surrounding, dehydrated medium tends to "suck out" the liquid remaining in the cell (see previous paragraph), causing bubble growth and eventually resulting in complete emptying of the cell (see Figure \ref{fig:BubbleDynamics}c); during this process the bubble tends to stay at the top of the cell due to buoyancy.

The timescales of emptying strongly depend on the volume of the cell, and of the transport properties and geometry of the surrounding medium. For experiments with hydrogels or synthetic xylem based on porous silicon (see section \ref{sec:CavitationPropagation}), the emptying time, $\tau_\mathrm{empt}$, is on the order of minutes, and can be predicted based on the permeability and poroelastic constants of the material \cite{Vincent2012,Vincent2014}.
In real xylem, values between tens of milliseconds and tens of minutes are predicted, depending on the type of tree \cite{Holtta2007}.
All these estimates are many orders of magnitude slower than the initial dynamics of the cavitation bubble leading to temporary equilibrium (see above), so that the assumption used in section \ref{sec:NucleationTheory} (i.e., considering that the number of water molecules stays constant in the cell during nucleation and bubble oscillations), is justified.

\subsection{Discussion}

Cavitation bubble dynamics in closed, cell-like environments displays rich features with timescales spanning from sub-microsecond to minutes for a single bubble (see Figure \ref{fig:BubbleDynamics}).
In the context of plants, some of these features are potentially interesting.
For example, the clear oscillatory signature could help analyze acoustic signals measured on dehydrating plants.
Also, cavitation close to surfaces (e.g. boat propellers) is usually associated with damage (due to strong, focused jets, shockwaves etc. when the bubble collapses on itself in the vicinity of the surface \cite{Brennen2014}).
We may hypothesize that such damaging phenomena are not present in plants, because cavitation bubbles, rather than collapsing, tend to grow and oscillate around a finite size.
One way to explain this difference is that bubbles in plants appear in a tissue that is dehydrated and under static negative pressure, while bubbles close to propellers appear due to a very transient metastable state due to flow, and tend to collapse back completely due to the positive pressure of the stagnant, surrounding fluid.
Another noteworthy remark is that the complex bubble dynamics makes it challenging to know the exact location where cavitation was initiated.
Indeed, because the bubble oscillates quickly and is then ejected from the walls within microseconds, erroneous conclusions on the initial bubble position can be easily drawn when not recording the dynamics with very large frame rates.

It has to be noted that the conclusions above are based on experiments in model systems with spherical geometry.
It is not well known yet how bubble dynamics changes when the shape differs from this ideal situation.
Cubic cells have shown a quasi-identical behavior as spherical cells in terms of the oscillatory dynamics \cite{Scognamiglio2018}, but it would be useful to investigate much more elongated structures such as those found in xylem.
From experiments in microfluidics with open channels \cite{Zwaan2007}, one could expect rich shape dynamics driven by the geometry of the confining cell, potentially very different from the spherical case.
The last stage of bubble dynamics (slow growth due to cell emptying, see Figure \ref{fig:BubbleDynamics}), however, probably depends less on the exact shape of the cells and more on the global transport properties in the whole tissue.
This expansion stage has important consequences for the general dynamics of propagation in multi-cell structures, as we will explain in the next section.

\section{Propagation of cavitation \label{sec:CavitationPropagation}}

So far we have explored the birth and growth of a cavitation bubble in a single cell.
Plant tissues contain many cells connected to one another and the way bubbles from one cell propagate (or not) to other cells has important consequences.
In the fern leptosporangium (see Figure \ref{fig:CavitationInPlants}c), the ejection mechanism requires cavitation bubbles to appear simultaneously in a maximum number of cells in the structure; it is thus favorable in such a situation to have a mechanism that allows fast propagation of cavitation through the tissue.
On the contrary, cavitation is detrimental in xylem because it results in embolism; there, propagation is unfavorable, and indeed observations in xylem show that cavitation occurs through discrete events (see Figure \ref{fig:CavitationInPlants}a-b).
Below, we will discuss phenomena that either trigger (positive interaction) or suppress (negative interaction) cavitation in neighboring cells. Surprisingly, both effects can result in bursts of cavitation events, but for different reasons and on different timescales.

\begin{figure*}[]
  \begin{center}
  \includegraphics[scale=0.75]{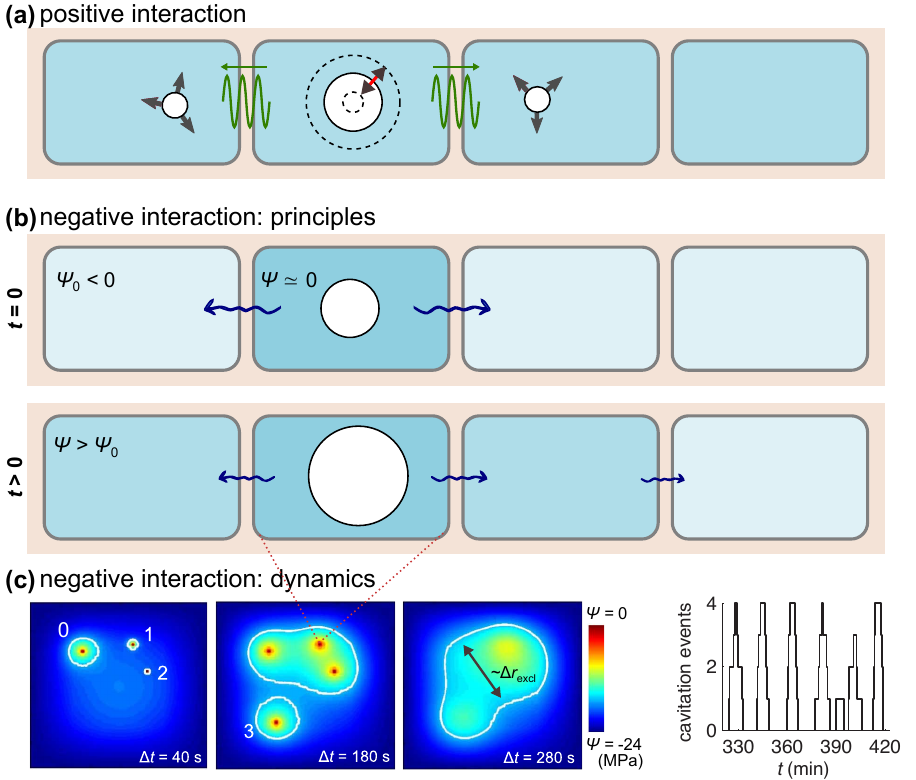}
  \caption{\small
  Propagation of cavitation in multi-cell systems, either due to positive interactions between cells (a) or negative interactions (b-c), see text for details.
  (a) The oscillation of a bubble in one cell creates a pressure disturbance sufficient to trigger cavitation in the neighboring cells.
  (b) Cavitation in one cell can also temporarily prevent cavitation in neighboring cells, because the cavitated cell acts as a source of water that rehydrates its surroundings while the cell is emptying (i.e. while the bubble is growing); $t=0$ refers to the appearance of the cavitation bubble, and colors indicate the evolution of pressure in each cell (darker corresponds to higher pressure).
  (c) As a result, cavitation events develop an \emph{exclusion zone} of larger pressure around them, where cavitation is hindered. The sequence on the left panel shows a simulation of the evolution of the pressure field due to 4 distinct cavitation events (labeled 0 to 3 in chronological order) in a multicellular structure;
  the white lines represent iso-pressure contours at $-17$ MPa, inside which cavitation probability is essentially zero.
  The panel on the right is an experimental recording of cavitation "bursts" resulting from the complex relaxation dynamics of the pressure field.
  Panel (c) reprinted with permission from \cite{Vincent2014}, \copyright 2014 by the
  American Physical Society."
  }
\label{fig:CavitationPropagation}
\end{center}
\end{figure*}

\subsection{Triggered cavitation: positive interactions}

Simultaneous cavitation in neighboring cells has been observed in the fern leptosporangium (see Figure \ref{fig:CavitationInPlants}c) and also in artificial, hydrogel-based biomimetic systems, either due to spontaneous, neighbor-to-neighbor propagation \cite{Pellegrin2015} or due to an externally induced shockwave that travels through all cells \cite{Vincent2012a}.
In this last situation, it is straightforward to understand the process: all cells initially contain liquid at negative pressure, and the shockwave passing through the liquid provides a short burst of energy that is sufficient to overcome the energy barrier of nucleation in the metastable liquid.
In the first case where simulataneous cavitation is spontaneous, there is no external signal triggering cavitation, but the disturbance associated with the stochastic nucleation of a cavitation bubble in any of the cells can be sufficient to trigger nucleation in neighboring cells (see Figure \ref{fig:CavitationPropagation}a).
Cavitation can then propagate from cell to cell in a chain-reaction fashion.
While not much is known about the conditions (cell-to cell separation, wall stiffness, etc.) required for cavitation to propagate in this manner, theoretical investigations have confirmed that cell-to-cell propagation is possible due to acoustic coupling between neighboring cells \cite{Doinikov2018,ElAmri2017a}.
Interestingly, the timescales of propagation are predicted to be strongly linked to the oscillation dynamics of the cavitation bubble: the strongest probability of triggering cavitation occurs in a neighboring cell around the collapse phase after 1 cycle of oscillation.
Since oscillation frequencies are typically in the MHz range (see section \ref{sec:BubbleDynamics}), neighbor-to-neighbor triggering should be almost instantaneous, which is consistent with experimental observations \cite{Noblin2012,Pellegrin2015}.
More experimental and modelling investigations are necessary to characterize these propagation phenomena.

\subsection{Hindered cavitation: negative interactions}

If the acoustic coupling between neighboring cells is not sufficiently strong, the positive triggering described in the previous section does not happen.
However the cells can communicate hydraulically if their walls (or pits in the case of xylem) allow for water flow.
Such flow is reponsible for the last stage of bubble dynamics, i.e. the slow growth of the bubble (cell emptying) over long timescales ($\sim$ minutes) (see section \ref{sec:BubbleDynamics});
its origin is the pressure imbalance between the relaxed liquid contained in the cell after cavitation ($P \simeq P_\mathrm{sat} > 0$) and the liquid in the surrounding cells, which is still at negative pressure.
This outward flow \emph{rehydrates} the neighboring cells, resulting in an increase of pressure in these cells, by a mechanism symmetrical to that described in section \ref{sec:DehydrationMechanics} (see Figure \ref{fig:CavitationPropagation}b).
This increase in pressure makes the neighboring cells less likely to cavitate (see section \ref{sec:NucleationTheory}), and thus cavitation in one cell tends to hinder cavitation in the immediate vicinity (negative interaction).

The pressure increase progressively spreads to other cells further away, forming an \emph{exclusion zone} around the initial cavitation event, where cavitation is suppressed (see Figure \ref{fig:CavitationPropagation}c, left).
The extent, $\Delta r$ of this zone grows over time, $t$, and its growth is dictated by the poroelastic transport properties in the medium : $\Delta r \sim \sqrt{C t}$ where $C \, \mathrm{[m^2 / s]}$ is an effective diffusivity sometimes called a consolidation coefficient. $C$ scales as the ratio of permeability of the structure to its effective compressibility, $\chi = \chi_\ell + \chi_\mathrm{c}$ \cite{Wang2017a,Vincent2014}, showing once again the importance of compressibility effects in cavitation dynamics.
The growth of the exclusion zone is fed by the emptying of the cell where cavitation initially occurred, and this feeding stops when the cell is empty.
As a result, the maximum extent of the exclusion zone is $\Delta r_\mathrm{excl} \sim \sqrt{C \tau_\mathrm{empt}}$, where $\tau_\mathrm{empt}$ is the emptying time (see section \ref{sec:BubbleDynamics});
$\Delta r_\mathrm{excl}$ can be seen as the radius of influence of a cavitation event in the medium.
After the cavitated cell has completely emptied, the associated exclusion zone progressively vanishes.
Consequently, the exclusion zone effect is both limited in space (radius of influence $\Delta r_\mathrm{excl}$) and time (duration on the order of $\tau_\mathrm{empt}$).

The dynamics described above of rehydration with the formation of exclusion zones around cavitation events was demonstrated experimentally in artificial systems based on porous silicon \cite{Vincent2014}.
An interesting consequence is that despite having \emph{negative} interactions between neighboring cells, the propagation of cavitation can occur by bursts of seemingly simultaneous cavitation events (see Figure \ref{fig:CavitationPropagation}c, right).
The reason for these bursts is the complex evolution of the pressure field in the system, associated with the regular formation and disappearance of multiple exclusion zones due to stochastic cavitation events.
This complex, nonlinear coupling between poroelastic transport and nucleation results in self-organized spatio-temporal patterns that create burst-like behavior, i.e. periods with several cavitation events followed by "quiet" periods with no cavitation events.
Compared to bursts originating from positive interactions (see previous section), the timescales involved here are much slower: in the artificial, xylem-like systems \cite{Vincent2014}, individual events within a burst could be separated by tens of seconds, while the bursts themselves happened every $15$ minutes approximately.

Potentially, in real xylem these timescales could be shorter or longer ($\sim$ hours) depending on the geometry of the cells and physical properties of the tissue (elasticity, permeability).
In a living tree, these cavitation/rehydration cycles could provide temporary relief in dry conditions by "sacrificing" some water-conducting cells to rehydrate the surrounding tissues.
These cycles could also be related to the puzzling observation that sometimes the sap flow rate and xylem water potential in real plant leaves can oscillate with periods on the order of the hour \cite{Lang1969,Steppe2006}.

\subsection{Discussion}

From the discussion above, it is obvious that the dynamics of cavitation in multi-cell systems is rich and complex, and poorly explored.
While a few experiments artificial systems have allowed us to unveil unexpected features of cavitation propagation, probing the dynamics of formation of bubbles in real systems is challenging, in particular for xylem in trees.
Non-invasive techniques such as magnetic resonance imaging or X-ray microtomography are powerful but limited in sample size and/or time resolution \cite{Cochard2013}.
Much simpler optical techniques have recently enabled direct visualization of embolism development in xylem slices \cite{Ponomarenko2014}, living leaves \cite{Brodribb2016a} or freshly cut branches \cite{Brodribb2017}; they are limited to samples sufficiently thin to allow for light propagation but are promising as a tool to investigate cavitation and embolism dynamics without the need of heavy experimental equipment.
Combined approaches using modelling, experiments with artificial systems and experiments on real plant tissues are needed to make progress on the understanding of cavitation dynamics and its implications.

\section*{Conclusion}

Plants live in an out-of equilibrium world where the atmosphere is often sub-saturated in water vapor (humidity lower than 100\%RH).
As a result, they naturally dehydrate, putting the liquid they contain into a state of negative pressure, i.e. mechanical tension.
Plants exploit this metastable state in various ways: in xylem, negative pressure is used as a pulling force to lift water from the roots to the leaves; in some ferns and fungi, cavitation in the metastable liquid provides a sudden release of energy for spore ejection strategies.
However, cavitation in xylem also results in embolism, which is detrimental to the plant.
Optimized compromises between efficiency and safety have thus evolved since the first plants appeared on land.

In this chapter, we have discussed the basic physics of formation and propagation of cavitation bubbles in cellular structures such as those found in plants and fungi.
As we have shown, this physics is rich, from the fluid mechanics of ultra-fast bubble oscillations to the complex spatio-temporal patterns of nucleation in systems with multiple cells in interaction.
Most of these results are support by experimental investigations on artificial systems mimicking some aspects of plants, however the dynamics of bubbles in real plants is far from being well understood.

The development of synthetic structures has two goals.
First, these structures can help understanding what is happening inside a real plant by studying samples that are much simpler, controlled and accessible than, e.g. a living tree.
On this aspect, current developments should probably aim at building structures that are closer to real plant tissues in terms of geometry, permeability, elasticity etc.
For example, there is currently some effort to understand bubble penetration through the nanoporous pits separating xylem vessels and e.g. the role of surfactants; artificial systems with pore sizes similar to those of pits (typically 10-100 nm) would probably prove useful in these investigations.
The synthetic structures based on porous silicon or hydrogels mentioned in this chapter currently have pores with diameters of only a few nanometers, which make bubble penetration extremely difficult and unlikely.

The second goal of artificial systems is to develop technologies inspired by plants to solve various problems in engineering, taking advantage of solutions already present in nature.
This direction has already proven fruitful, with examples of biomimetic systems using negative pressure to pump water \cite{Wheeler2008,Shi2020}, transport heat \cite{Chen2014a}, or desalinate water \cite{Wang2020}.
Improving these prototypes, but also explore other applications such as energy harvesting or mechanical actuation, should provide large inspiration for scientists in the near future.

\section*{Acknowledgments}

I am thankful to Teemu H\"oltt\"a, Benjamin Dollet, Yo\"el Forterre, Steven Jansen, Pono and others who have provided useful information during the preparation of this chapter through discussions and/or e-mail exchanges.
I also thank Philippe Marmottant, Abraham Stroock, students and postdocs in their groups, and other colleagues with whom I have worked on the themes of negative pressure and cavitation in various contexts.

\appendix

\section{Effect of air on the saturation vapor pressure of water \label{sec:AppendixAirVaporPressure}}

Equilibrium between liquid water and pure water vapor (in the absence of air) occurs when the two phases are at the saturation pressure, $P_\mathrm{sat}(T)$, by definition. In this situation, the chemical potentials are equal in both phases ($\mu_\ell(P_\mathrm{sat}) = \mu_\mathrm{v}(P_\mathrm{sat}) \equiv \mu_\mathrm{sat}(T)$).

In the presence of air, the liquid and vapor phases are at a different pressure: due to mechanical equilibrium with air, liquid water is at atmospheric pressure, $P = P_\mathrm{atm}$, while water vapor is characterized by its partial vapor pressure in air, $p$.
By integration from the pure substance equilibrium ($\mu_\mathrm{sat}$), the corresponding chemical potentials are
\begin{equation}
    \mu_\ell = \mu_\mathrm{sat} + v_\mathrm{m} (P - P_\mathrm{sat})
    \label{eq:MuLiq}
\end{equation}
for liquid water (assuming an incompressible liquid), and
\begin{equation}
  \mu_\mathrm{v} = \mu_\mathrm{sat} + RT \ln \left( \frac{p}{P_\mathrm{sat}} \right)
  \label{eq:MuVap}
\end{equation}
for water vapor, assuming that it behaves as an ideal gas. Derivation of equations (\ref{eq:MuLiq}-\ref{eq:MuVap}) can be done using the thermodynamic relation equating the partial derivative of $\mu$ with respect to pressure to the molar volume, $v$, of the substance (constant $v = v_\mathrm{m}$ for the liquid phase, $v = RT / p$ for the vapor phase).

We define $p_\mathrm{sat}$ as the equilibrium partial vapor pressure, $p$ when the liquid is maintained at atmospheric pressure, $P = P_\mathrm{atm}$. Using the equality of chemical potentials between Equations (\ref{eq:MuLiq}-\ref{eq:MuVap}),
\begin{equation}
    p_\mathrm{sat} = P_\mathrm{sat} \exp \left( \frac{v_\mathrm{m} (P_\mathrm{atm} - P_\mathrm{sat})}{RT} \right)
\end{equation}
which yields $p_\mathrm{sat} / P_\mathrm{sat} = 1.0007$ at $25 \degree C$; this factor depends weakly on temperature.

At first sight, one could think that because $p_\mathrm{sat} > P_\mathrm{sat}$, water vapor at $p_\mathrm{sat}$ would be supersaturated, i.e. metastable with respect to condensation. However this is not the case, because condensation is also affected by air pressure: the condensed liquid phase forms at $P_\mathrm{atm}$, not at $P_\mathrm{sat}$. This extra pressure makes condensation only favorable when $p > p_\mathrm{sat}$. As a result, the equilibrium with liquid water at $P_\mathrm{atm}$ and water vapor at partial pressure $p_\mathrm{sat}$ in air is a true, stable equilibrium shifted from $P_\mathrm{sat}$ because of air pressure.
Using this shifted equilibrium as the new reference state ($\mu_\mathrm{sat}^\mathrm{air}$), the chemical potentials can be rewritten
\begin{align}
  \begin{cases}
      \mu_\ell & = \mu_\mathrm{sat}^\mathrm{air} + v_\mathrm{m} (P - P_\mathrm{atm}) \\
      \mu_\mathrm{v} & = \mu_\mathrm{sat}^\mathrm{air} + RT \ln \left( \frac{p}{p_\mathrm{sat}} \right)
      \label{eq:MuLiqVapAir}
  \end{cases}
\end{align}
for the liquid and vapor phases, respectively.

We finally note that air has another potential effect on the liquid-vapor equilibrium of water. Indeed, by dissolving in liquid water at a concentration $C_\mathrm{air}$, air results in an osmotic pressure $\Pi_\mathrm{air} \simeq RTC_\mathrm{air}$ with an associated shift in chemical potential $\Delta \mu_\mathrm{air,sol} = -v_\mathrm{m} \Pi_\mathrm{air}$. From the solubility of air in water \cite{Sander2015}, one can estimate $\Pi_\mathrm{air} \simeq 2$ kPa when the water is saturated with air. Since $\Pi_\mathrm{air} \ll P_\mathrm{atm} \simeq 100$ kPa, $\Delta \mu_\mathrm{air}^\mathrm{sol} = -v_\mathrm{m} \Pi_\mathrm{air}$ is negligible compared to the shift in chemical potential introduced by air pressure, $\Delta \mu_\mathrm{air}^\mathrm{pressure}\simeq v_\mathrm{m} P_\mathrm{atm}$.
In other words, the effect of dissolved air on the saturation vapor pressure is orders of magnitude smaller than the effect of the mechanical pressure of air, which is itself already quite small.

\section{Free energy of a confined bubble \label{sec:AppendixNucleationTheory}}

Here, we derive the free energy landscape of a vapor bubble (volume, $V$) in a liquid ($V_\ell$), both enclosed in a container ($V_\mathrm{c}$) that is elastic (potentially, infinitely stiff). Thus,
\begin{equation}
V_\mathrm{c} = V + V_\ell.
\label{eq:Volumes}
\end{equation}
We use the state of the system at $V=0$ (homogeneous liquid, no bubble) as the reference state, where the liquid pressure is $P_0$ (see Figure \ref{fig:ConfinedNucleation}).
Since the liquid occupies the whole container, $V_\ell = V_\mathrm{c}$ in this reference state;
we define the reference volume, $V_0$ as being the value of $V_\ell$ and $V_\mathrm{c}$ in this situation.
Note that the reference pressure $P_0$ is not atmospheric pressure here, but the initial pressure of the liquid, which is negative in the situations of interest in this chapter.

We consider an isothermal transformation, with a constant total number of water molecules (liquid + vapor) in the system.
This assumption is justified by the fact that during the timescales of interest for bubble nucleation, the liquid does not have time to flow out of the cell (see section \ref{sec:CellEmptying}).
Pressure in the liquid varies as a function of bubble size: bubble expansion compresses the liquid ($V_\ell$ decreases) and makes its pressure $P$ increase.
This increase in pressure also makes the container expand ($V_\mathrm{c}$ increases if not infinitely stiff), and as we will see below, we can use knowledge about the variations of $V_\mathrm{c}$ with $P$ to fully characterize the free energy variations of the \emph{outside world} due to variations in bubble size.

For such a system, the natural thermodynamic potential is the Helmholtz free energy, $F$, and equilibrium is obtained when $F$ is minimal.
We use the notation $\Delta F = F - F_0$ to describe the Helmholtz free energy difference between the current state (bubble volume, $V$) and the reference state ($V=0$).

Below we evaluate the different contributions to the total Helmholtz free energy variation $\Delta F$ of the system as a function of bubble size ($V$): liquid pressure ($P$) and compressibility ($\chi_\ell$), elastic deformation of the container ($\chi_\mathrm{c}$), surface tension of the liquid-vapor interface ($\gamma$), and evaporation of the liquid into the bubble (vapor pressure, $P_\mathrm{sat}$).
The derivation below follows a different and more general approach compared to earlier versions \cite{Vincent2012a, Vincent2017a}, but yields the same final result.

\paragraph{Liquid pressure and compressibility}

We use the linearized equation of state of liquid water (Equation \ref{eq:Compressibility}), taking the homogeneous liquid at negative pressure as the reference state:
\begin{equation}
    V_\ell - V_\mathrm{0} = - \chi_\ell V_0 (P - P_0).
    \label{eq:LiquidEquationOfState}
\end{equation}
We first consider that the liquid does not evaporate into the bubble and thus evolves at constant substance amount $n$ when the bubble forms.
In a second step, we will calculate the correction due to evaporation (see \emph{Evaporation into bubble} paragraph below).
Because $dF = -S \mathrm{d}T - P \mathrm{d}V + \mu \mathrm{d}n$ for a pure substance (with parameters $n, V, T$), then $\Delta F = - \int P \mathrm{d}V$ for constant $T, n$ and we can directly obtain by integration of Equation (\ref{eq:LiquidEquationOfState}) the free energy of the liquid as a function of its volume:
\begin{equation}
    \Delta F_\ell = - P_0 (V_\ell - V_0) + \frac{1}{2} \frac{1}{\chi_\ell V_0} \left( V_\ell - V_0 \right)^2.
    \label{eq:FreeEnergyLiquid}
\end{equation}

\paragraph{Evaporation into bubble}

In reality, a small fraction of the liquid evaporates into the bubble (amount $\Delta n$). We make the simplifying assumption that this evaporation maintains the vapor in the bubble at the saturated vapor pressure, $P_\mathrm{sat}$.
There are subtleties in the consequences of this assumption but is impact on the results in classical nucleation theory is negligible \cite{Blander1975}.

Since $F = \mu n - P V$, the total free energy of the vapor in the bubble is
\begin{equation}
  F_\mathrm{vap} = \mu_\mathrm{sat} \Delta n - P_\mathrm{sat} V
  \label{eq:EvapVapor}
\end{equation}
where $\mu_\mathrm{sat}$ is the chemical potential of saturated vapor, which is also by definition the chemical potential of liquid water at pressure $P_\mathrm{sat}$, because liquid and vapor are in equilibrium when $P = P_\mathrm{sat}$.

In our situation, however, the liquid is not at $P_\mathrm{sat}$ but at $P$, so its chemical potential is $\mu_\mathrm{sat} + v_\mathrm{m} (P - P_\mathrm{sat})$, neglecting second-order compressibility effects.
Thus, using $F = \mu n - P V$ again, the change in free energy of the liquid due to evaporation is $\Delta F_\mathrm{liq,evap} = -\Delta n (\mu_\mathrm{sat} + v_\mathrm{m}(P - P_\mathrm{sat})) - P (-v_\mathrm{m} \Delta n)$, or
\begin{equation}
  \Delta F_\mathrm{liq,evap} = \mu_\mathrm{sat} (- \Delta n) + v_\mathrm{m} \Delta n P_\mathrm{sat}
  \label{eq:EvapLiquid}
\end{equation}

As a result, the total effect of evaporation is the sum of the contributions of Equations (\ref{eq:EvapVapor}) and (\ref{eq:EvapLiquid})
\begin{equation}
  \Delta F_\mathrm{evap} = P_\mathrm{sat} \left( v_\mathrm{m} \Delta n - V \right) \simeq - P_\mathrm{sat} V
  \label{eq:FreeEnergyEvaporation}
\end{equation}
where we neglect $v_\mathrm{m} \Delta n$ compared to $V$ because of the large difference between liquid and vapor densities; the term $v_\mathrm{m} \Delta n$ should be kept when working close to the critical point, but we do not consider such situations here.

Note that it is also possible to add contributions from other gases present in the bubble \cite{Vincent2017a}, but we neglect this effect here. In particular, we assume that dissolved air does not have time to significantly fill the bubble during the timescales of nucleation.

\paragraph{Elasticity of container}

When pressure changes in the liquid, the container has an elastic response that we characterize using a linear approximation similar to the one we used for the compressibility of water (see section \ref{sec:Compressibility}):
\begin{equation}
  V_\mathrm{c} - V_\mathrm{0} = \chi_\mathrm{c} V_0 (P - P_0).
  \label{eq:ContainerEquationOfState}
\end{equation}
For example, if the container is a spherical inclusion in an infinite, incompressible solid, $\chi_\mathrm{c} = 3 / (4 G)$ where $G$ is the shear modulus of the medium \cite{Vincent2017a}.

When $V_\mathrm{c}$ changes by $\mathrm{d} V_\mathrm{c}$, the \emph{outside world} (including the container itself) changes volume by $-\mathrm{d} V_\mathrm{c}$ and receives work from the fluid (pressure $P$) by an amount $\int P \mathrm{d} V_\mathrm{c}$. By definition, its free energy changes by the same amount, so that the contribution of the container deformation (and of the outside world) to the changes in the total free energy can be calculated from integration of Equation (\ref{eq:ContainerEquationOfState})
\begin{equation}
    \Delta F_\mathrm{c} = P_0 (V_\mathrm{c} - V_0) + \frac{1}{2} \frac{1}{\chi_\mathrm{c} V_0} \left( V_\mathrm{c} - V_0 \right)^2.
    \label{eq:FreeEnergyContainer}
\end{equation}
Note the similarity but also the differences in sign compared to Equation (\ref{eq:FreeEnergyLiquid}), due to the fact that while liquid contracts when pressure increases, the container expands.
Note also that we do not need to know the details of the physical processes occurring in the outside world that result in how much the container deforms under pressure (e.g. entropic vs mechanical, etc.): only knowing the relationship between container volume and inner pressure (Equation \ref{eq:ContainerEquationOfState}) is sufficient to characterize the outside world contribution to the total free energy.

\paragraph{Total free energy}

We can now write the total Helmholtz free energy $\Delta F$ as a function of the volumes $V$, $V_\ell$, $V_\mathrm{c}$ by summing all the contributions above (Equations \ref{eq:FreeEnergyLiquid}, \ref{eq:FreeEnergyEvaporation}, \ref{eq:FreeEnergyContainer}) and the contribution of the surface tension $\gamma$ of the liquid-vapor interface, which by definition is  $\gamma A$, where $A$ is the surface area of the bubble:
\begin{equation}
    \small
    \Delta F = \gamma A + (P_0 - P_\mathrm{sat}) V + \frac{1}{2 V_0} \left[ \frac{1}{\chi_\ell} (V_\ell - V_0)^2 + \frac{1}{\chi_\mathrm{c}} (V_\mathrm{c} - V_0)^2\right]
    \label{eq:FreeEnergyTotalRaw}
\end{equation}
where we have used Equation (\ref{eq:Volumes}) to simplify some terms.

We can further simplify Equation (\ref{eq:FreeEnergyTotalRaw}) by using the relationship between $V$, $V_\mathrm{c}$ and $V_\ell$ that is implied by the liquid and container equations of state. Indeed, from Equations (\ref{eq:LiquidEquationOfState}) and (\ref{eq:ContainerEquationOfState}), and using $V = V_\mathrm{c} - V_\ell$, it is straightforward to show that
\begin{align}
\begin{cases}
    V_\ell - V_0 & = - \frac{\chi_\ell}{\chi_\ell + \chi_\mathrm{c}} V \\
    V_\mathrm{c} - V_0 & = \frac{\chi_\mathrm{c}}{\chi_\ell + \chi_\mathrm{c}} V \\
    P - P_0 & = \frac{1}{\chi_\ell + \chi_\mathrm{c}} \frac{1}{V_0} V
    \label{eq:VolumesPressureVolumeBubble}
\end{cases}
\end{align}
These equations can also be found by direct minimization of $\Delta F$ with respect to $V_\ell$ and $V_\mathrm{c}$ for an imposed value of $V$.
Injecting the volume relations from Equation (\ref{eq:VolumesPressureVolumeBubble}) into Equation (\ref{eq:FreeEnergyTotalRaw}) yields
\begin{equation}
    \Delta F = \gamma A + (P_0 - P_\mathrm{sat}) V + \frac{1}{2} \frac{1}{V_0 (\chi_\ell + \chi_\mathrm{c})} V^2.
    \label{eq:FreeEnergyTotal}
\end{equation}
Note that we have not made assumptions about the shape of the bubble or container, and our calculations thus apply to arbitrary shapes.

We also note that Equation (\ref{eq:FreeEnergyTotal}) is less general than Equation (\ref{eq:FreeEnergyTotalRaw}) because the former uses the equilibrium relations (\ref{eq:VolumesPressureVolumeBubble}) and thus implicitly assumes that liquid and container volumes evolve in mechanical equilibrium given a constraint of bubble volume of size $V$. In other words, Equation (\ref{eq:FreeEnergyTotal}) already uses a partial minimization of Equation (\ref{eq:FreeEnergyTotalRaw}) to relate $V_\ell$ and $V_\mathrm{c}$ to $V$ and reduce the number of degrees of freedom to one ($V$). By contrast, Equation (\ref{eq:FreeEnergyTotalRaw}) has two degrees of freedom (three variables $V_\mathrm{c}$, $V_\ell$, $V$ with the constraint $V + V_\ell = V_\mathrm{c}$) and could in principle be used to describe situations where these quantities evolve independently (e.g. oscillation of the container and bubble that are not in phase). This may be confusing at first because we initially used equilibrium relations to derive the free energies of the liquid and of the elastic deformation of the container. However, this is only an illustration of the fact that reversible transformations can be used to probe the state variables of thermodynamic systems.

\bibliography{References}

\end{document}